\documentclass{article}
\usepackage{amsfonts}
\usepackage{makeidx}
\usepackage{latexsym,amsmath,amssymb,amscd}
\usepackage{makecell}
\usepackage{xcolor}
\usepackage[all]{xy}
\usepackage{endnotes}

\let\footnote=\endnote

\newtheorem{theorem}{Theorem}

\newtheorem{proposition}[theorem]{Proposition}

\allowdisplaybreaks

\begin{document}

\title{First Integrals of holonomic systems without Noether symmetries}
\author{Michael Tsamparlis$^{1,a)}$ and Antonios Mitsopoulos$^{1,b)}$ \\
%EndAName
{\ \ }\\
$^{1}${\textit{Faculty of Physics, Department of
Astronomy-Astrophysics-Mechanics,}}\\
{\ \textit{University of Athens, Panepistemiopolis, Athens 157 83, Greece}}
\vspace{12pt} %EndAName
\\
$^{a)}$Email: mtsampa@phys.uoa.gr %EndAName
\\
$^{b)}$Author to whom correspondence should be addressed: antmits@phys.uoa.gr }
\date{}
\maketitle

\begin{abstract}
A theorem is proved which determines the first integrals
of the form $I=K_{ab}(t,q)\dot{q}^{a}\dot{q}^{b}+K_{a}(t,q)\dot{q}^{a}+K(t,q)$
of autonomous  holonomic systems using only the collineations of the kinetic
metric which is defined by the kinetic energy or the Lagrangian of the system. It is
shown how these first integrals can be associated via the inverse Noether
theorem to a gauged weak Noether symmetry which admits the given first integral as a
Noether integral. It is shown also that the associated Noether symmetry is possible to satisfy the
conditions for a Hojman or a form-invariance symmetry therefore the so-called non-Noetherian
first integrals are gauged weak Noether integrals.
 The application of the theorem
requires a certain algorithm due to the complexity of the special conditions involved. We
demonstrate this algorithm by a number of solved examples. We choose examples from published
works in order to show that our approach produces new first integrals not found before
with the standard methods.
\end{abstract}

\section{Introduction}

The symmetries of dynamical systems, described by differential equations,
are described/understood in two different ways.

a. The geometric method. In this method one identifies the differential
equation with the set of all the solution curves. Then a symmetry of the
differential equation is understood as a point transformation in the
appropriate space of the variables which preserves the set of solution
curves in the sense, that takes one point of a solution curve to a point on
another solution curve. The point transformation is defined by means of a
vector field which is called the generator of the transformation. If the
point transformation depends only on the variables and not on their
derivatives it is just called a point transformation and the generator is a
vector field in the (base) space defined by the variables of the equation.
If the point transformation contains derivatives of the dependent variables
it is called a generalized transformation (contact transformations, B\"{a}%
cklund transformations etc.) whereas the generator of the transformation is
a vector field on the appropriate jet space over the base space. Finally if
the solution curves are parameterized then the point transformation is
possible to preserve or not the parameter and it is characterized
accordingly.

b. The algebraic method. In this method the symmetry of the equation is a
change of coordinates in the space of the variables so that in the new
variables the resulting differential equation has the same form as the
original, therefore it has the same set of solutions. It is possible that in
the new variables the symmetries of the differential equation are evident or
they can be understood. If this is the case, they are computed in the new
variables and then by the inverse transformation one gets them in the
original variables.

The geometric method is the prevailing one in practice. It was initiated and
systematized by Sophus Lie\footnote{S. Lie, ``Theorie der Transformationsgruppen I'', Leipzig: B.G. Teubner (1888).\label{lie1}}$^{,}$\footnote{S. Lie, ``Theorie der Transformationsgruppen II'', Leipzig: B.G. Teubner (1890).\label{lie2}}$^{,}$\footnote{S. Lie, ``Theorie der Transformationsgruppen III'', Leipzig: B.G. Teubner (1893).\label{lie3}} who used the theory of
continuous transformation groups to define the Lie symmetries. Using the Lie
symmetries one may define appropriate variables in which the differential
equation is reduced and in general it is simplified. Furthermore it is
possible that a Lie symmetry leads to a first integral (FI).

In the present work we consider the autonomous holonomic dynamical system of
the form
\begin{equation}
\ddot{q}^{a}=\omega ^{a}(q)  \label{FLII.0}
\end{equation}%
where $\omega ^{a}=-\Gamma _{bc}^{a}\dot{q}^{b}\dot{q}^{c}-V^{,a}(q)+F^{a}(q,%
\dot{q})$. In (\ref{FLII.0}) $\Gamma _{bc}^{a}$ are the Riemann  connection
coefficients of the metric   $\gamma _{ab}=\frac{\partial ^{2}L}{\partial
\dot{q}^{a}\partial \dot{q}^{b}}$ defined by the Lagrangian (or the kinetic
energy), $V$ stands for all conservative forces, $F^{a}$ for the
non-conservative generalized forces and the Einstein summation convention is
used. A dot indicates total derivative wrt the parameter $t$. Every such equation
defines the Hamiltonian vector field
\begin{equation*}
\mathbf{\Gamma }\equiv \frac{d}{dt}=\frac{\partial }{\partial t}+\dot{q}^{a}%
\frac{\partial }{\partial q^{a}}+\omega ^{a}\frac{\partial }{\partial \dot{q}%
^{a}}.
\end{equation*}%
The condition that the vector field
\begin{equation}
\mathbf{X}=\xi (t,q,\dot{q})\partial _{t}+\eta ^{a}(t,q,\dot{q})\partial
_{q^{a}}  \label{FLII.0.1}
\end{equation}%
is a Lie symmetry of (\ref{FLII.0}) is that there exists a function $\lambda
\left( t,q,\dot{q}\right) $ such that
\begin{equation}
\lbrack \mathbf{X}^{[1]},\mathbf{\Gamma }]=\lambda (t,q,\dot{q})\mathbf{%
\Gamma }  \label{FLII.0.2}
\end{equation}%
where
\begin{equation}
\mathbf{X}^{[1]}=\xi (t,q,\dot{q})\partial _{t}+\eta ^{a}(t,q,\dot{q}%
)\partial _{q^{a}}+\left( \dot{\eta}^{a}-\dot{q}^{a}\dot{\xi}\right)
\partial _{\dot{q}^{a}}  \label{FLII.0.1.1}
\end{equation}%
is the first prolongation\footnote{%
This is the complete lift of $\mathbf{X}$ in $TM$.} of $\mathbf{X}$ in the first jet
space $J^{1}(t,q,\dot{q})$. It can be shown that the Lie symmetry condition (%
\ref{FLII.0.2}) can be written
\begin{equation}
\mathbf{X}^{[2]}H^{a}=0\implies \mathbf{X}^{[1]}H^{a}+\eta ^{a[2]}=0
\label{FLII.0.3}
\end{equation}%
where $H^{a}\equiv \ddot{q}^{a}-\omega ^{a}$, $\eta ^{a[2]}=\ddot{\eta}^{a}-2%
\ddot{q}^{i}\dot{\xi}-\dot{q}^{i}\ddot{\xi}$ and $\mathbf{X}^{[2]}=\mathbf{X}%
^{[1]}+\eta ^{a[2]}\partial _{\ddot{q}^{a}}$ is the second prolongation of $%
\mathbf{X}$ in $J^{2}\left\{ t,q^{a},\dot{q}^{a},\ddot{q}^{a}\right\} $.

As it is well-known\footnote{H. Stephani, ``Differential Equations: Their Solutions using Symmetry'', Cambridge University Press, New York (1989).\label{StephaniB}} a velocity-dependent Lie
symmetry has an extra degree of freedom which allows the introduction of a
scalar condition, in which case one speaks for a gauged Lie symmetry. The
standard gauge condition is $\xi=0 $ which simplifies the Lie symmetry
condition (\ref{FLII.0.3}) as follows
\begin{equation}
\mathbf{X}^{[1]}(\omega ^{a})= \mathbf{\Gamma}\left( \mathbf{\Gamma}%
(\eta^{a}) \right) = \ddot{\eta}^{a}.  \label{FLII.0.4}
\end{equation}

\subsection{The case of First Integrals (FIs)}

FIs are the most useful tool in the study of the dynamical equations. The first major
contribution  on this topic is the work of Noether\footnote{E. Noether, Transp. Theory Statist. Phys. \textbf{1}(3), 186 (1971). \label{Noether1}}
who introduced the Noether symmetries. A Noether symmetry of a Lagrangian
system is a Lie symmetry which satisfies in addition the Noether condition.
Geometrically a Noether symmetry is a point transformation\footnote{%
The original work of Noether considers the general space $J^{k}\left( t,q,%
\dot{q},...,q^{(k)}\right) $ where $k=1,2,...$ and $q^{(k)}\equiv \frac{%
d^{k}q}{dt^{k}}$. In this work we shall be working on the $J^{1}(t,q,\dot{q})
$ because we are interested in QFIs. For a recent discussion of the Noether
approach see the following: A.K. Hadler, A. Paliathanasis and
P.G.L. Leach, Symmetry \textbf{10}(12), 744 (2018).\label{Hadler Paliathanasis Leach 2018}} in $J^{1}(t,q,\dot{q})$
under which the action integral is invariant up to a perfect differential
with zero endpoint variation, so that the resulting Euler-Lagrange
equations (hence the set of solutions) remain the same. The important result
is that to each Noether symmetry there corresponds a concomitant first
integral, called a Noether integral.

The Noether integrals are usually autonomous but they can also be time-dependent. The time-dependent FIs are equally important as the autonomous
ones. Indeed  time-dependent FIs can be used to test the (Liouville)
integrability and also the superintegrability of a Hamiltonian system.
Specifically, the Liouville theorem\footnote{See p. 271, Sec. 49 in V.I. Arnold, ``Mathematical Methods of Classical Mechanics'', Springer, (1989), proof in p. 272-284.\label{Arnold 1989}} on integrability requires $n$ functionally independent FIs (i.e.
their gradients on the phase space are linearly independent) in involution of the form $I(q,p)$. However, one can also use time-dependent FIs\footnote{See Theorem 1, p.17, Chap. II,
Para 2 in V.V. Kozlov, Russ. Math. Surv., Turpion, \textbf{38}(1), pp. 1-76 (1983). \label{Kozlov 1983}}$^{,}$\footnote{See Theorem 3.4 in T.G. Vozmishcheva, J. Math. Sc. \textbf{125}(4), 419 (2005). \label{Vozmishcheva 2005}} of the form $I(q,p,t)$. It is to be
noticed that both Theorems in Refs. \ref{Kozlov 1983} and \ref{Vozmishcheva 2005} refer to a time-dependent
Hamiltonian $H(q,p,t)$.

Perhaps the next most systematic important approach after the Noetherian one
is the direct approach of Katzin and Levine, originated by Darboux,\footnote{G. Darboux, Archives Neerlandaises (II) \textbf{6}, 371 (1901). \label{Darboux}} which is discussed in section \ref{sec.lie.FIs.collineations}.

An additional different approach has been developed by Hojman\footnote{S.A. Hojman, J. Math. Phys. A: Math. Gen. \textbf{25}, L291 (1992). \label{Hojman 1992}} who showed that under certain conditions a Lie symmetry leads to a FI,
called a Hojman integral. These FIs are coordinate-dependent therefore they
are not useful (at least in Physics where the Covariance Principle requires
that the physical quantities must be covariant wrt the fundamental group of
the theory). Finally it has also been shown that under certain conditions a
form-invariance symmetry\footnote{%
In the case of a holonomic dynamical system $E_{a}(L)=F_{a}$ where $F_{a}$
are the non-conservative generalized forces and $E_{a}=\frac{d}{dt}\frac{%
\partial }{\partial \dot{q}^{a}}-\frac{\partial }{\partial q^{a}}$ is the
Euler-Lagrange vector field, the form-invariance symmetry satisfies the
condition
\begin{equation*}
E_{a}\left( \Delta L(t,q,\dot{q})\right) =\Delta F_{a}(t,q,\dot{q})\implies
E_{a}\left( \mathbf{X}^{[1]}(L)\right) =\mathbf{X}^{[1]}\left( F_{a}\right) .
\end{equation*}%
} is also possible to give a FI.\footnote{F.X. Mei, H.B. Wu and Y.F. Zhang, Int. J. Dynam. Control \textbf{2}, 285 (2014). \label{Mei et all}}

In conclusion, there are only two important and systematic approaches for the
computation of the FIs: The Noether approach and the direct approach of
Katzin and Levine.

\section{The conditions for a weak Noether symmetry}

For holonomic dynamical systems defined by a Lagrangian $L(t,q,\dot{q})$ and
generalized non-conservative forces $F^{a}$ one has to use the weak Noether condition\footnote{D.S. Djukic and B.D. Vujanovic, Acta Mechanica \textbf{23}, 17-27 (1975). \label{Djukic1975}}
\begin{equation}
\mathbf{X}^{W}(L) + L\dot{\xi} =\mathbf{X}^{\left[ 1\right] }\left( L\right) +\phi ^{a}\frac{\partial L}{%
\partial \dot{q}^{a}}+L\dot{\xi}=\dot{f}  \label{GenHolonNoether'sFinCond}
\end{equation}%
where the function $f(t,q^{a},\dot{q}^{a})$ is called the Noether or the
gauge function and to the vector field (\ref{FLII.0.1}) corresponds the weak
first prolongation
\begin{equation}
\mathbf{X}^{W}\equiv \mathbf{X}^{[1]}+\phi ^{a}(t,q,\dot{q})\frac{\partial }{%
\partial \dot{q}^{a}}.  \label{eq.weak.1}
\end{equation}%
The weak Noether condition leads to the FI
\begin{equation}
I=f-L\xi -\frac{\partial L}{\partial \dot{q}^{a}}\left( \eta ^{a}-\xi \dot{q}%
^{a}\right)   \label{GenHolonNoether's1Integr}
\end{equation}%
provided that the functions $\phi ^{a}(t,q,\dot{q})$ are defined by the
condition
\begin{equation}
F_{a}\left( \eta ^{a}-\xi \dot{q}^{a}\right) =\phi ^{a}\frac{\partial L}{%
\partial \dot{q}^{a}}.  \label{GenHolonNoether's2Cond}
\end{equation}%
Substituting (\ref{GenHolonNoether's2Cond}) in (\ref%
{GenHolonNoether'sFinCond}) we obtain the so-called Noether-Bessel-Hagen
(NBH) equation
\begin{equation}
\mathbf{X}^{\left[ 1\right] }\left( L\right) +L\dot{\xi}+F_{a}\left( \eta
^{a}-\xi \dot{q}^{a}\right) =\dot{f}.  \label{NBH}
\end{equation}

For velocity-dependent Noether symmetries in the gauge $\xi =0$ the latter
conditions simplify as follows%
\begin{equation}
I=f-\frac{\partial L}{\partial \dot{q}^{a}}\eta ^{a}
\label{GenHolonNoether's1Integr2}
\end{equation}
\begin{equation}
F_{a}\eta ^{a}=\phi ^{a}\frac{\partial L}{\partial \dot{q}^{a}}.
\label{GenHolonNoether's2Cond2}
\end{equation}

The FIs of Hojman and the ones defined by the form-invariance symmetry have
been called non-Noetherian FIs because the generators of the corresponding
point transformations do not satisfy the weak Noether condition.

However there is an alternative approach to look at the non-Noetherian and
the Noetherian FIs. Indeed according to the Inverse Noether
Theorem to every FI one may associate  (in general) a velocity-dependent
gauged Noether symmetry whose generator is not necessarily the
same with the one deriving the non-Noetherian FI. Therefore, in
a sense, all FIs are or can be Noether integrals. The Inverse
Noether theorem for velocity-dependent Noether symmetries has as follows (see Ref. \ref{Djukic1975}).

\begin{theorem}
\label{Inverse Noether Theorem} \textbf{(Inverse Noether theorem)} Suppose $\Lambda$ is a FI of a holonomic dynamical
system with regular Lagrangian $L(t,q^{a},\dot{q}^{a})$ and generalized non-conservative forces $F^{a}(t,q,\dot{q})$. Then the vector $\mathbf{X}=\xi(t,q,\dot{q})\frac{\partial}{\partial t} + \eta^{a}(t,q,\dot{q})\frac{\partial}{\partial q^{a}}$ with a weak first prolongation
\begin{equation}
\mathbf{X}^{W}= \mathbf{X}^{[1]} + \phi^{a}(t,q,\dot{q})\partial _{\dot{q}^{a}} = \xi\partial _{t}+\eta^{a}\partial_{q^{a}} +\left( \dot{\eta}^{a}-\dot{q}^{a}\dot{\xi} + \phi^{a}\right)
\partial_{\dot{q}^{a}}   \label{Prf13.3}
\end{equation}%
is the generator of a weak Noether
symmetry with gauge function $f(t,q,\dot{q})$ provided that
\begin{eqnarray}
\eta^{a} &=& -\gamma^{ab} \frac{\partial \Lambda}{\partial \dot{q}^{b}} + \xi \dot{q}^{a} \label{eq.InvNoe.1a} \\
\phi^{a} \frac{\partial L}{\partial \dot{q}^{a}} &=& -F^{a} \frac{\partial \Lambda}{\partial \dot{q}^{a}} \label{eq.InvNoe.1b} \\
\xi &=& \frac{1}{L} \left( f - \Lambda + \gamma^{ab} \frac{\partial L}{\partial \dot{q}^{a}} \frac{\partial \Lambda}{\partial \dot{q}^{b}} \right). \label{eq.InvNoe.1c}
\end{eqnarray}
This weak Noether symmetry produces the given FI $\Lambda$. Therefore any FI for such systems can be associated to a weak Noether symmetry.
\end{theorem}

\textbf{Proof.}
\bigskip

We write the Euler-Lagrange equations as follows
\begin{equation}
\ddot{q}^{a} = \gamma^{ab} \left( F_{b} + \frac{\partial L}{\partial q^{b}} - \frac{\partial^{2} L}{\partial t\partial \dot{q}^{b}} - \frac{\partial^{2} L}{\partial \dot{q}^{b} \partial q^{c}} \dot{q}^{c} \right) \label{eq.InvNoe.2}
\end{equation}
where $\gamma_{ab}\equiv \frac{\partial^{2} L}{\partial \dot{q}^{a} \partial \dot{q}^{b}}$. Using (\ref{eq.InvNoe.2}) the FI condition $\frac{d\Lambda}{dt}=0$ gives
\begin{equation}
\frac{\partial \Lambda}{\partial t} + \frac{\partial \Lambda}{\partial q^{a}} \dot{q}^{a} + \gamma^{ab} \frac{\partial \Lambda}{\partial \dot{q}^{a}} \left( F_{b} + \frac{\partial L}{\partial q^{b}} - \frac{\partial^{2} L}{\partial t\partial \dot{q}^{b}} - \frac{\partial^{2} L}{\partial \dot{q}^{b} \partial q^{c}} \dot{q}^{c} \right) =0. \label{eq.InvNoe.3}
\end{equation}

Taking into account the above results it is sufficient to show that the weak Noether condition (\ref{GenHolonNoether'sFinCond}) is satisfied for the set $(\xi, \eta^{a}, \phi^{a}, f)$ defined by the conditions (\ref{eq.InvNoe.1a}) - (\ref{eq.InvNoe.1c}). We compute
\begin{eqnarray*}
\dot{\xi} &=& - L^{-2} \left( \frac{\partial L}{\partial t} + \frac{\partial L}{\partial q^{c}}\dot{q}^{c} + \frac{\partial L}{\partial \dot{q}^{c}} \ddot{q}^{c} \right) \left( f - \Lambda +\gamma^{ab} \frac{\partial L}{\partial \dot{q}^{a}} \frac{\partial \Lambda}{\partial \dot{q}^{b}} \right) + \\
&&+L^{-1} \left[ \dot{f} + \dot{\gamma}^{ab} \frac{\partial L}{\partial \dot{q}^{a}} \frac{\partial \Lambda}{\partial \dot{q}^{b}} + \gamma^{ab} \frac{d}{dt} \left(\frac{\partial L}{\partial \dot{q}^{a}}\right) \frac{\partial \Lambda}{\partial \dot{q}^{b}} + \gamma^{ab} \frac{\partial L}{\partial \dot{q}^{a}}\frac{d}{dt} \left( \frac{\partial \Lambda}{\partial \dot{q}^{b}}\right) \right] \\
&=& - L^{-2} \left( \frac{\partial L}{\partial t} + \frac{\partial L}{\partial q^{c}}\dot{q}^{c} + \frac{\partial L}{\partial \dot{q}^{c}} \ddot{q}^{c} \right) \left( f - \Lambda +\gamma^{ab} \frac{\partial L}{\partial \dot{q}^{a}} \frac{\partial \Lambda}{\partial \dot{q}^{b}} \right) + \\
&& + L^{-1} \left[ \dot{f} + \dot{\gamma}^{ab} \frac{\partial L}{\partial \dot{q}^{a}} \frac{\partial \Lambda}{\partial \dot{q}^{b}} +F^{a}\frac{\partial \Lambda}{\partial \dot{q}^{a}} +\gamma^{ab} \frac{\partial L}{\partial q^{a}} \frac{\partial \Lambda}{\partial \dot{q}^{b}} + \gamma^{ab} \frac{\partial L}{\partial \dot{q}^{a}}\frac{d}{dt} \left( \frac{\partial \Lambda}{\partial \dot{q}^{b}} \right) \right]
\end{eqnarray*}
and
\begin{eqnarray*}
\dot{\eta}^{a} &=& -\dot{\gamma}^{ab} \frac{\partial \Lambda}{\partial \dot{q}^{b}} -\gamma^{ab} \left( \frac{\partial^{2} \Lambda}{\partial t \partial \dot{q}^{b}} + \frac{\partial^{2} \Lambda}{\partial \dot{q}^{b} \partial q^{c}} \dot{q}^{c} + \frac{\partial^{2} \Lambda}{\partial \dot{q}^{b} \partial \dot{q}^{c}} \ddot{q}^{c} \right) + \dot{\xi} \dot{q}^{a} + \xi \ddot{q}^{a}.
\end{eqnarray*}
Substituting $\xi,\eta^{a}$ from (\ref{eq.InvNoe.1a}) - (\ref{eq.InvNoe.1c}) and the  total derivatives computed above in (\ref{GenHolonNoether'sFinCond}) we find that the weak Noether condition is trivially satisfied. Therefore the set $(\xi, \eta^{a}, \phi^{a}, f)$ generates a weak Noether symmetry whose FI is the FI $I=\Lambda$. This completes the proof.
\bigskip

In the case of the gauge $\xi=0$ the conditions  defining a gauged weak Noether (generalized) symmetry are reduced as follows
\begin{eqnarray}
\eta^{a} &=& -\gamma^{ab} \frac{\partial \Lambda}{\partial \dot{q}^{b}} \label{eq.InvNoe.g1a} \\
\phi^{a} \frac{\partial L}{\partial \dot{q}^{a}} &=& -F^{a} \frac{\partial \Lambda}{\partial \dot{q}^{a}} \label{eq.InvNoe.g1b} \\
f &=& \Lambda + \eta^{a}\frac{\partial L}{\partial \dot{q}^{a}}. \label{eq.InvNoe.g1c}
\end{eqnarray}

Moreover, by applying the Inverse Noether Theorem to a general QFI of the form
\[
\Lambda = K_{ab}(t,q)\dot{q}^{a}\dot{q}^{b} + K_{a}(t,q)\dot{q}^{a} + K(t,q)
\]
we deduce from the conditions (\ref{eq.InvNoe.g1a}) - (\ref{eq.InvNoe.g1c}) that $\Lambda$ is associated to the gauged weak Noether symmetry as follows
\begin{eqnarray}
\eta^{a} &=& -\gamma^{ab} \left( 2K_{bc}\dot{q}^{c} +K_{b} \right) \label{eq.InvNoe.g2a} \\
\phi^{a} \frac{\partial L}{\partial \dot{q}^{a}} &=& -F^{a} \left( 2K_{ab}\dot{q}^{b} +K_{a} \right) \label{eq.InvNoe.g2b} \\
f &=& K_{ab}\dot{q}^{a}\dot{q}^{b} + K_{a}\dot{q}^{a} + K -\gamma^{ab} \left( 2K_{bc}\dot{q}^{c} +K_{b} \right) \frac{\partial L}{\partial \dot{q}^{a}} \label{eq.InvNoe.g2c}
\end{eqnarray}
where $\frac{\partial \Lambda}{\partial \dot{q}^{a}}= 2K_{ab}\dot{q}^{b} +K_{a}$.

For $L= \frac{1}{2} \gamma_{ab} \dot{q}^{a} \dot{q}^{b} - V(q)$ the conditions (\ref{eq.InvNoe.g2a}) - (\ref{eq.InvNoe.g2c}) become
\begin{eqnarray}
\eta^{a} &=& -\gamma^{ab} \left( 2K_{bc}\dot{q}^{c} +K_{b} \right) \label{eq.InvNoe.g2na} \\
\left(\phi_{a} +2K_{ab}F^{b} \right)\dot{q}^{a} + K_{a}F^{a} &=& 0 \label{eq.InvNoe.g2nb} \\
f &=& -K_{ab}\dot{q}^{a}\dot{q}^{b} + K. \label{eq.InvNoe.g2nc}
\end{eqnarray}

It is easy to check that the gauged weak Noether symmetry
\begin{equation}
\left( \xi=0, \eta_{a} = -2K_{ab}\dot{q}^{b} -K_{a}, \phi_{a}, f= -K_{ab}\dot{q}^{a}\dot{q}^{b} + K \right) \enskip \text{such that $\left(\phi_{a} +2K_{ab}F^{b} \right)\dot{q}^{a} + K_{a}F^{a} =0$} \label{eq.GaugeNoether1}
\end{equation}
does produce the Noether FI $\Lambda$. Another gauged weak Noether symmetry which generates the same result is the
\begin{equation}
\left( \xi=0, \eta_{a}= -K_{ab}\dot{q}^{b} -K_{a}, \phi_{a}, f=K \right) \enskip \text{such that $\left(\phi_{a} +K_{ab}F^{b} \right)\dot{q}^{a} + K_{a}F^{a} =0$.} \label{eq.GaugeNoether2}
\end{equation}

\section{Lie symmetries, FIs and collineations}

\label{sec.lie.FIs.collineations}

Over the years various works have appeared with the view to put the Lie and
the Noether approach in geometric terms. The reason for this is that if one
manages to relate the Lie and the Noether symmetries with the symmetries
(collineations) of the kinetic metric, which is defined by the dynamical system
itself, then one may use the vast results of Differential
Geometry to compute the generators of the Lie/Noether symmetries admitted by the
dynamical system and consequently the FIs. It appears that the first clear
approach in this direction was done in the 80s mainly by Katzin and
Levine.\footnote{G.H. Katzin and J. Levine, J. Math. Phys. \textbf{9}(1), 8 (1968). \label{Katzin 1968}}$^{,}$\footnote{G.H. Katzin, J. Math. Phys. \textbf{14}(9), 1213 (1973).\label{Katzin 1973}}$^{,}$\footnote{G.H. Katzin and J. Levine, J. Math. Phys. \textbf{15}(9), 1460 (1974). \label{Katzin 1974}}$^{,}$\footnote{G.H. Katzin and J. Levine, J. Math. Phys. \textbf{17}(7), 1345 (1976). \label{Katzin 1976}}$^{,}$\footnote{G.H. Katzin and J. Levine, J. Math. Phys. \textbf{22}(9), 1878 (1981). \label{Katzin 1981}}$^{,}$\footnote{G.H. Katzin and J. Levine, J. Math. Phys. \textbf{26}(12), 3080 (1985). \label{KatzinLev1985}}

These authors considered a holonomic conservative system and showed that the
generators of the Lie point symmetries are the projective collineations
(PCs) of the kinetic metric and the generators of the Noether point
symmetries are the elements of the homothetic algebra of the kinetic metric.
Perhaps these results were not emphasized enough and due to
the special $\delta$-derivative approach they used, it appears that they passed
rather unnoticed. As a result other works followed using the same approach\footnote{T.M. Kalotas and B.G. Wybourne, J. Phys A: Math. Gen. \textbf{15}, 2077 (1982). \label{Kalotas}}$^{,}$\footnote{H. Kaplan, Am. J. Phys. \textbf{54}, 157 (1986). \label{Kaplan}} concerning mainly special types of autonomous conservative systems (see also Refs \ref{StephaniB} and \ref{Djukic1975}). The results of these latter works follow as special cases of the general results of Katzin and Levine.
A clear, complete and systematic presentation\footnote{M. Tsamparlis and A. Paliathanasis, J. Phys. A: Math. Theor. \textbf{44}, 175202 (2011). \label{Tsamp 2011}} of this approach has been
given recently in Ref. \ref{Tsamp 2011} which has lead to a number of
interesting applications.\footnote{T. Sen, Phys. Lett. A \textbf{122}(6,7), 327 (1987). \label{Sen}}

The approach of Katzin and Levine can be summarized as follows. Instead they to consider the Lie/Noether symmetries, which are the intermediate
steps in the determination of the FIs, they focused directly on
the computation of the FIs $I$ from the condition $\frac{dI}{dt}=0.$ To do
that they considered the quadratic FIs (QFIs) $I$ of the form
\begin{equation}
I=K_{ab}(t,q)\dot{q}^{a}\dot{q}^{b}+K_{a}(t,q)\dot{q}^{a}+K(t,q)
\label{FL.5}
\end{equation}%
where $K_{ab}$ is a symmetric tensor, $K_{a}$ is a vector and $K$ is an
invariant and required that $\mathbf{\Gamma}(I)=\frac{dI}{dt}=0.$ This requirement
leads to a system of conditions for the coefficients $K_{ab},K_{a},K$ whose
solution gives all second (and first)\ order autonomous and time-dependent
QFIs admitted by the dynamical equations. In this approach the problem of
determining the QFIs of dynamical equations is reduced
to the solution of the resulting system of equations for the coefficients $K_{ab},K_{a},K.$ Obviously these equations depend on the form of the general quantity $\omega^{a}$ (i.e. the particular dynamical system) whereas the solution of the resulting system is a formidable task.

In a recent paper\footnote{M. Tsamparlis and A. Mitsopoulos, J. Math. Phys. \textbf{61}, 072703 (2020). \label{Tsamp2020}} (henceforth referred as paper A) the
complete systematic solution of the system of these equations for the case of autonomous conservative dynamical systems has been given
under the assumption that the tensors involved have the general form $%
K_{ab}(t,q)=g(t)C_{ab}(q),$ $K_{a}(t,q)=f(t)L_{a}(q) +B_{a}(q)$, where $g(t), f(t)$ are analytic functions. This solution (Theorem 1 of paper A) provided all
the autonomous and the time-dependent QFIs of the dynamical equations (\ref%
{FLII.0}) for $F^{a}=0$ in terms of the collineations (including the Killing tensors-KTs) of
the kinetic metric.

The purpose of the present work is to generalize the results of paper A to
the case of autonomous holonomic dynamical systems which move in a
Riemannian space under the action of generalized forces of the form $%
F^{a}=-P^{a}(q)+A_{b}^{a}(q)\dot{q}^{b}$. The dynamical equations for these
systems are%
\begin{equation}
\ddot{q}^{a}=-\Gamma _{bc}^{a}\dot{q}^{b}\dot{q}^{c}-Q^{a}(q)+A_{b}^{a}(q)%
\dot{q}^{b}  \label{FL.5.1}
\end{equation}%
where the generalized forces $Q^{a}\equiv V^{,a}+P^{a}$ contain all the forces
conservative and non-conservative. We assume again that $I$ has the general
form (\ref{FL.5}) and determine the system of equations resulting from the
condition $dI/dt=0$ together with the integrability conditions for the
scalar $K.$ We solve this system of equations in terms of the collineations
of the kinetic metric (including the KTs) and the result is stated  in
Theorem \ref{Theorem2}.

The structure of the paper is as follows. In section \ref{section.1} we
derive the system of equations which result from the condition $dI/dt=0$ for
the dynamical system (\ref{FL.5.1}). These equations reduce to the corresponding equations of paper A for $F^{a}=0$. In section \ref{section.2} we give
the general solution of the system of equations as Theorem \ref{Theorem2}. In section \ref{sec.In} we study the behavior of the case Integral 1 of Theorem \ref{Theorem2}.
In section \ref{section.3} we give a brief theory concerning the
determination of KTs in a Riemannian space in terms of the collineations of
the kinetic metric. In sections \ref{sec.E2.geometry}, \ref{sec.KTE3} we present some useful results concerning the KTs of $E^{2}$ and $E^{3}$ which shall be used widely in applications of Theorem \ref{Theorem2} over Newtonian systems. In section \ref{applications} we demonstrate the significance of
Theorem \ref{Theorem2} by considering various examples. It is shown that
Theorem \ref{Theorem2} besides the Noether FIs also computes the
Hojman integrals and the form-invariance integrals. Finally in section \ref{section.5} we draw our conclusions. In Appendix we sketch the proof of
Theorem \ref{Theorem2}.

\section{The conditions for a QFI}

\label{section.1}

Equations (\ref{FLII.0}) may be considered as the Euler-Lagrange equations
for the Lagrangian $L(q,\dot{q})=\frac{1}{2}\gamma _{ab}\dot{q}^{a}\dot{q}%
^{b}-V(q)$ with generalized forces $F_{a}.$ The Lagrangian is assumed to be
regular, that is $det\frac{\partial ^{2}L}{\partial \dot{q}^{a}\partial \dot{%
q}^{b}}\neq 0,$ and defines the new non-degenerate kinetic metric $\gamma
_{ab}=$ $\frac{\partial ^{2}L}{\partial \dot{q}^{a}\partial \dot{q}^{b}}$.
The kinetic metric is not the metric of the space where motion occurs except
in the case of a free system (that is $V=0, F^{a}=0$) in which case equations
(\ref{FLII.0})\ are the geodesic equations. In the following the covariant
derivatives and the rising/lowering of indices are done with the kinetic
metric $\gamma _{ab}.$

We consider the function $I$ given by (\ref{FL.5}) and using the dynamical
equations (\ref{FL.5.1}) to replace the terms $\ddot{q}^{a}$ we write the condition $\frac{dI}{dt}=0$ as
\begin{align*}
0& =K_{(ab;c)}\dot{q}^{a}\dot{q}^{b}\dot{q}^{c}+\left(
K_{ab,t}+K_{a;b}+2K_{c(a}A_{b)}^{c}\right) \dot{q}^{a}\dot{q}^{b}+\left(
K_{a,t}+K_{,a}-2K_{ab}Q^{b}+\right. \\
& \quad \left. +K_{b}A_{a}^{b}\right) \dot{q}^{a}+K_{,t}-K_{a}Q^{a}
\end{align*}
from which follows the system of equations
\begin{eqnarray}
K_{(ab;c)} &=&0  \label{eq.veldep4.1} \\
K_{ab,t}+K_{(a;b)}+2K_{c(a}A_{b)}^{c} &=&0  \label{eq.veldep4.2} \\
-2K_{ab}Q^{b}+K_{a,t}+K_{,a}+K_{b}A_{a}^{b} &=&0  \label{eq.veldep4.3} \\
K_{,t}-K_{a}Q^{a} &=&0.  \label{eq.veldep4.4}
\end{eqnarray}
Condition (\ref{eq.veldep4.1}) implies that $K_{ab}$ is a Killing tensor
(KT) of order 2 (possibly zero) of the kinetic metric $\gamma _{ab}$.

The most general choice for the KT $K_{ab}$ in the case of an autonomous
system is\footnote{%
Equivalently we may assume $K_{ab}= \sum^{n}_{N=1} f_{N}(t) D_{(N)ab}(q)$
where $f_{N}(t)$ is a sequence of analytic functions and $D_{(N)ab}$ is a
sequence of KTs of $\gamma_{ab}$. This expression is equivalent to (\ref%
{eq.aspm1}) because if we write
\begin{equation*}
f_{N}(t)= \sum^{n}_{M=0} d_{(N)M} t^{M} = d_{(N)0} + d_{(N)1}t + ... +
d_{(N)n} t^{n}
\end{equation*}
then
\begin{equation*}
K_{ab}= \sum^{n}_{N=1} \sum^{n}_{M=0} d_{(N)M} t^{M} D_{(N)ab}(q) =
\sum^{n}_{M=0} \underbrace{\left( \sum^{n}_{N=1} d_{(N)M} D_{(N)ab}(q)
\right)}_{\equiv \bar{D}_{(M)ab}(q)} t^{M}= \sum^{n}_{M=0} \bar{D}%
_{(M)ab}(q) t^{M}.
\end{equation*}%
}
\begin{equation}
K_{ab}(t,q)=C_{(0)ab}(q) + \sum_{N=1}^{n}C_{(N)ab}(q)\frac{t^{N}}{N}
\label{eq.aspm1}
\end{equation}%
where $C_{(N)ab}$, $N=0,1,...,n$ is a sequence of arbitrary KTs of order 2
of the kinetic metric $\gamma_{ab}$.

This choice of $K_{ab}$ and equation (\ref{eq.veldep4.2}) indicate that we
set
\begin{equation}
K_{a}(t,q)=\sum_{M=0}^{m}L_{(M)a}(q)t^{M}  \label{eq.aspm2}
\end{equation}%
where $L_{(M)a}(q)$ are arbitrary vectors.

We note that both powers $n$, $m$ in the above polynomial expressions may be
infinite.

Substituting (\ref{eq.aspm1}), (\ref{eq.aspm2}) in the system of equations (%
\ref{eq.veldep4.1}) -(\ref{eq.veldep4.4}) (equation (\ref{eq.veldep4.1}) is
identically zero since $C_{(N)ab}$ are assumed to be KTs) we obtain
\begin{eqnarray}
&&  \notag  \label{eq.veldep5} \\
0
&=&C_{(1)ab}+C_{(2)ab}t+...+C_{(n)ab}t^{n-1}+L_{(0)(a;b)}+L_{(1)(a;b)}t+...+L_{(m)(a;b)}t^{m}+2C_{(0)c(a}A_{b)}^{c}+
\notag \\
&&+2C_{(1)c(a}A_{b)}^{c}t+...+2C_{(n)c(a}A_{b)}^{c}\frac{t^{n}}{n}
\label{eq.veldep6} \\
0 &=&-2C_{(0)ab}Q^{b}-2C_{(1)ab}Q^{b}t-...-2C_{(n)ab}Q^{b}\frac{t^{n}}{n}%
+L_{(1)a}+2L_{(2)a}t+...+mL_{(m)a}t^{m-1}+K_{,a}+L_{(0)b}A_{a}^{b}+  \notag
\\
&&+L_{(1)b}A_{a}^{b}t+...+L_{(m)b}A_{a}^{b}t^{m}  \label{eq.veldep7} \\
0 &=&K_{,t}-L_{(0)a}Q^{a}-L_{(1)a}Q^{a}t-...-L_{(m)a}Q^{a}t^{m}.
\label{eq.veldep8}
\end{eqnarray}

Conditions (\ref{eq.veldep5}) - (\ref{eq.veldep8}) must be supplemented with
the integrability conditions $K_{,at}=K_{,ta}$ and $K_{,[ab]}=0$ for the
scalar function $K$. The integrability condition $K_{,at}=K_{,ta}$ gives -
if we make use of (\ref{eq.veldep7}) and (\ref{eq.veldep8}) - the equation
\begin{eqnarray}
0 &=&-2C_{(1)ab}Q^{b}-2C_{(2)ab}Q^{b}t... -2C_{(n)ab}Q^{b}t^{n-1}+
2L_{(2)a}+6L_{(3)a}t+...+m(m-1)L_{(m)a}t^{m-2}+\left( L_{(0)b}Q^{b}\right)
_{,a}+  \notag \\
&&+\left( L_{(1)b}Q^{b}\right)_{,a}t +...+\left( L_{(m)b}Q^{b}\right)
_{,a}t^{m} + L_{(1)b}A_{a}^{b} +2L_{(2)b}A_{a}^{b}t
+...+mL_{(m)b}A_{a}^{b}t^{m-1}.  \label{eq.veldep9}
\end{eqnarray}%
Condition $K_{,[ab]}=0$ gives the equation
\begin{eqnarray}
0 &=&2\left( C_{(0)[a\left\vert c\right\vert }Q^{c}\right) _{;b]}+2\left(
C_{(1)[a\left\vert c\right\vert }Q^{c}\right) _{;b]}t+
...+2\left(C_{(n)[a\left\vert c\right\vert }Q^{c}\right) _{;b]}\frac{t^{n}}{n%
} - L_{(1)\left[a;b\right] } -2L_{(2)\left[ a;b\right] }t-...-mL_{(m)\left[
a;b\right] }t^{m-1}-  \notag \\
&&-L_{(0)c;[b}A_{a]}^{c}-L_{(1)c;[b}A_{a]}^{c}t-...
-L_{(m)c;[b}A_{a]}^{c}t^{m} - L_{(0)c}A_{[a;b]}^{c}
-L_{(1)c}A_{[a;b]}^{c}t-... -L_{(m)c}A_{[a;b]}^{c}t^{m}  \label{eq.veldep10}
\end{eqnarray}%
which for 2d systems with $F^{a}=0$ is known as the second order
Bertrand-Darboux equation (see Ref. \ref{Darboux}).

Equations (\ref{eq.veldep6}) - (\ref{eq.veldep10}) constitute the system of
equations we have to solve.

\section{The Theorem}

\label{section.2}

The solution of the system of equations (\ref{eq.veldep6}) - (\ref%
{eq.veldep10}) can be found in the Appendix and it is stated in Theorem \ref%
{Theorem2} below.

\begin{theorem}
\label{Theorem2} The independent QFIs of a dynamical system  (\ref%
{FL.5.1}) are the following\footnote{%
We note that the FI $J_{1}$ is for $n$ finite whereas $J_{2}$ is for $n$
infinite hence the term $e^{\lambda t}$.}: \bigskip

\textbf{Integral 1.}
\begin{eqnarray*}
J_{1} &=&\left( \frac{t^{n}}{n}C_{(n)ab}+...+\frac{t^{2}}{2}%
C_{(2)ab}+tC_{(1)ab}+C_{(0)ab}\right) \dot{q}^{a}\dot{q}^{b}+t^{n}L_{(n)a}
\dot{q}^{a}+ ... +t^{2}L_{(2)a}\dot{q}^{a}+tL_{(1)a}\dot{q}^{a} +L_{(0)a}%
\dot{q}^{a}+ \\
&&+\frac{t^{n+1}}{n+1}L_{(n)a}Q^{a}+...+\frac{t^{2}}{2}%
L_{(1)a}Q^{a}+tL_{(0)a}Q^{a}+G(q)
\end{eqnarray*}%
where\footnote{%
We note that for $n=0$ the conditions for the QFI $J_{1}(n=0)$ can be
derived if we set equal to zero the quantities $C_{(N)ab}$ and $L_{(N)a}$
for $N\neq 0$.} $C_{(N)ab}$ for $N=0,1,...,n$ are KTs, $%
C_{(1)ab}=-L_{(0)(a;b)}-2C_{(0)c(a}A_{b)}^{c}$, $C_{(k+1)ab}=-L_{(k)(a;b)}-%
\frac{2}{k}C_{(k)c(a}A_{b)}^{c}$ for $k=1,...,n-1$, $L_{(n)(a;b)}=-\frac{2}{n%
}C_{(n)c(a}A_{b)}^{c}$, $\left( L_{(k-1)b}Q^{b}\right)
_{,a}=2C_{(k)ab}Q^{b}-k(k+1)L_{(k+1)a}-kL_{(k)b}A_{a}^{b}$ for $k=1,...,n-1$%
, $\left( L_{(n-1)b}Q^{b}\right) _{,a}=2C_{(n)ab}Q^{b}-nL_{(n)b}A_{a}^{b}$, $%
L_{(n)a}Q^{a}=s$ and $G_{,a}=2C_{(0)ab}Q^{b}-L_{(1)a}-L_{(0)b}A_{a}^{b}$.

\textbf{Integral 2.}
\begin{equation*}
J_{2}= e^{\lambda t} \left( \lambda C_{ab} \dot{q}^{a} \dot{q}^{b} + \lambda
L_{a}\dot{q}^{a} + L_{a}Q^{a} \right)
\end{equation*}
where $\lambda \neq 0$, $C_{ab}$ is a KT, $\lambda C_{ab} = - L_{(a;b)} -
2C_{c(a} A^{c}_{b)}$ and $\left(L_{b}Q^{b}\right)_{,a} = 2\lambda C_{ab}
Q^{b} - \lambda^{2}L_{a} - \lambda L_{b}A^{b}_{a}$.
\end{theorem}

We note that in all cases $C_{(N)ab}$ are KTs of order two whereas in many
cases the vector $K^{a}$ is a KV. This emphasizes the already known result
from previous studies (see Refs. \ref{StephaniB}, \ref{Kalotas} and \ref{Katzin 1981}) of the important role played by the KTs  and the KVs of the kinetic metric
in the determination of the FIs of (\ref{FL.5.1}).

In the case\footnote{%
If in addition $F^{a}=0$ the $Q^{a}=V^{,a}$  the case is reduced
to that of the autonomous conservative systems.} $A_{b}^{a}(q)=0$ Theorem \ref%
{Theorem2} takes the following form.

\begin{theorem}
\label{The first integrals of an autonomous holonomic dynamical system} The
independent QFIs of the dynamical system (\ref{FL.5.1}) for $A^{a}_{b}=0$
are the following: \bigskip

\textbf{Integral 1.}
\begin{eqnarray*}
I_{(1)} &=& \left( - \frac{t^{2\ell}}{2\ell} L_{(2\ell-1)(a;b)} - ... - \frac{%
t^{4}}{4} L_{(3)(a;b)} - \frac{t^{2}}{2} L_{(1)(a;b)} + C_{ab} \right) \dot{q%
}^{a} \dot{q}^{b} + t^{2\ell-1} L_{(2\ell-1)a}\dot{q}^{a} + ... +
t^{3}L_{(3)a}\dot{q}^{a} + \\
&& + t L_{(1)a}\dot{q}^{a} + \frac{t^{2\ell}}{2\ell} L_{(2\ell-1)a}Q^{a} +
... + \frac{t^{4}}{4} L_{(3)a}Q^{a} + \frac{t^{2}}{2} L_{(1)a}Q^{a} + G(q)
\end{eqnarray*}
where\footnote{%
We note that for $\ell=0$ the conditions for the QFI $I_{(1)}(\ell=0)$ are
given by nullifying all the vectors $L_{(M)a}$.} $C_{ab}$, $L_{(M)(a;b)}$
for $M=1,3,...,2\ell-1$ are KTs, $\left( L_{(2\ell-1)b} Q^{b} \right)_{,a} =
-2L_{(2\ell-1)(a;b)}Q^{b}$, $\left( L_{(k-1)b} Q^{b} \right)_{,a} =
-2L_{(k-1)(a;b)}Q^{b} - k(k+1)L_{(k+1)a}$ for $k=2,4,...,2\ell-2$ and $%
G_{,a}= 2C_{ab}Q^{b} - L_{(1)a}$.

\textbf{Integral 2.}
\begin{eqnarray*}
I_{(2)} &=& \left( - \frac{t^{2\ell+1}}{2\ell+1} L_{(2\ell)(a;b)} - ... -
\frac{t^{3}}{3} L_{(2)(a;b)} - t L_{(0)(a;b)} \right) \dot{q}^{a} \dot{q}%
^{b} + t^{2\ell} L_{(2\ell)a}\dot{q}^{a} + ... + t^{2}L_{(2)a}\dot{q}^{a} +
\\
&& + L_{(0)a}\dot{q}^{a}+ \frac{t^{2\ell+1}}{2\ell+1} L_{(2\ell)a}Q^{a} +
... + \frac{t^{3}}{3} L_{(2)a}Q^{a} +t L_{(0)a}Q^{a}
\end{eqnarray*}
where $L_{M(a;b)}$ for $M=0,2,...,2\ell$ are KTs, $\left( L_{(2\ell)b} Q^{b}
\right)_{,a} = -2L_{(2\ell)(a;b)}Q^{b}$ and $\left( L_{(k-1)b}
Q^{b}\right)_{,a} = -2L_{(k-1)(a;b)}Q^{b} - k(k+1)L_{(k+1)a}$ for $%
k=1,3,...,2\ell-1$.

\textbf{Integral 3.}
\begin{equation*}
I_{(3)} = e^{\lambda t} \left(-L_{(a;b)}\dot{q}^{a}\dot{q}^{b} + \lambda L_{a} \dot{q}^{a} + L_{a}Q^{a} \right)
\end{equation*}
where $L_{a}$ is such that $L_{(a;b)}$ is a KT and $\left(L_{b}Q^{b}
\right)_{,a} = -2L_{(a;b)} Q^{b} - \lambda^{2} L_{a}$.
\end{theorem}

We observe that for $A_{b}^{a}=0$ the QFI $J_{1}$ breaks  into two
independent QFIs the $I_{(1)}$ and $I_{(2)}$ corresponding to  even and odd powers of
$t.$ The case of autonomous conservative dynamical systems is obtained if
one sets $Q^{a}=V^{,a}$. Theorem \ref{The first integrals of an autonomous
holonomic dynamical system} is a generalized version of Theorem 1 of paper A because the assumption (\ref{eq.aspm1}) is more general than the one made in paper A.

It is apparent that before one attempts to compute the QFIs of a given
dynamical system of the form (\ref{FL.5.1}) using Theorem \ref{Theorem2} one
has to know the collineations of the kinetic metric including the second
order KTs. This is not a trivial requirement. However because the kinetic
metric is non-degenerate (the Lagrangian is assumed to be regular)\ it is
always possible to bring it to its canonical form by means of a proper
change of the coordinates and then use existing results of Differential
Geometry to compute the collineations and its KTs. A particular important
and fairly general case is that of spaces of constant curvature where these
quantities are known.\footnote{G. Thompson, J. Math. Phys. \textbf{27}(11), 2693 (1986). \label{Thompson1986A}}. However in general one has
to use special methods to compute the KTs.\footnote{P. Sommers, J. Math. Phys. \textbf{14}(6), 787 (1973). \label{Sommers 1973}}$^{,}$\footnote{E.G. Kalnins and W. Miller Jr., SIAM J. Math. Anal. \textbf{11}(6), 1011 (1980). \label{Kalnins 1980}}$^{,}$\footnote{R. Rani, S.B. Edgar and A. Barnes, Class. Quant. Grav. \textbf{20}, 1929 (2003). \label{Rani 2003}}$^{,}$\footnote{B. Coll, J.J. Ferrando and J.A. S\'{a}ez, J. Math. Phys. \textbf{47}, 062503 (2006). \label{Coll 2006}}$^{,}$\footnote{M. Crampin, Rep. Math. Phys. \textbf{62}(2), 241 (2008). \label{Crampin 2008}}$^{,}$\footnote{D. Garfinkle and E.N. Glass, Class. Quant. Grav. \textbf{27}, 095004 (2010). \label{Glass Garfinkle 2010}}

In section \ref{section.3} we consider briefly the determination of KTs from
the projective collineations (PCs) in a Riemannian space and state the
results for the case of spaces of constant curvature.

\section{Computing $J_{1}\equiv I_{n}$ in terms of the $I_{0}$}

\label{sec.In}

We prove that all QFIs $I_{N}$ where $N=1,2,...,n$ of the case \textbf{Integral 1} of Theorem \ref{Theorem2} can be constructed from the QFI $I_{0}$ by using the following systematic algorithm.
\bigskip

1) Write the QFI $I_{0}$.\newline

2) Introduce a new KT $C_{(1)ab}$ and a new vector $L_{(1)a}$. \newline

3) Construct $I_{1}$ by adding to the expression $I_{0}$ the time-dependent
terms $tC_{(1)ab}\dot{q}^{a}\dot{q}^{b}$, $tL_{(1)a}\dot{q}^{a}$ and $\frac{%
t^{2}}{2}L_{(1)a}Q^{a}$. \newline

4) Expand the conditions for $I_{0}$ so as to satisfy the requirement $\frac{%
dI_{1}}{dt}=0$. \newline

5) Continue in a similar manner with the construction of $I_{2}$ by using $%
I_{1}$. \newline

6) After some steps use the $I_{n-1}$ to construct $I_{n}$ by adding the
terms $\frac{t^{n}}{n} C_{(n)ab}\dot{q}^{a} \dot{q}^{b}$, $t^{n}L_{(n)a}\dot{%
q}^{a}$ and $\frac{t^{n+1}}{n+1} L_{(n)a}Q^{a}$.
\bigskip

We illustrate the above procedure for the small values of $n$.
\bigskip

- For $n=0$:

We have the QFI
\begin{equation*}
I_{0}=C_{(0)ab}\dot{q}^{a}\dot{q}^{b}+L_{(0)a}\dot{q}^{a}+ st +G(q)
\end{equation*}%
where $C_{(0)ab}$ is a KT and $L_{(0)a},G$ are computed from the expressions
\begin{equation*}
L_{(0)(a;b)}=-2C_{(0)c(a}A_{b)}^{c},\enskip L_{(0)b}Q^{b}=s,\enskip %
G_{,a}=2C_{(0)ab}Q^{b}-L_{(0)b}A_{a}^{b}.
\end{equation*}

- For $n=1$.

We have the QFI
\begin{equation*}
I_{1}=\left( tC_{(1)ab}+C_{(0)ab}\right) \dot{q}^{a}\dot{q}%
^{b}+tL_{(1)a}\dot{q}^{a}+L_{(0)a}\dot{q}^{a}%
+\frac{t^{2}}{2}s+tL_{(0)a}Q^{a}+G(q)
\end{equation*}%
where $C_{(1)ab}$ is a KT computed from the relation
\begin{equation*}
C_{(1)ab}=-L_{(0)(a;b)}-2C_{(0)c(a}A_{b)}^{c}
\end{equation*}%
and the vector $L_{(1)a}$ and the quantity $G$ are computed from the
relations
\begin{equation*}
L_{(1)(a;b)}=-2C_{(1)c(a}A_{b)}^{c},\enskip L_{(1)a}Q^{a}=s,\enskip\left(
L_{(0)b}Q^{b}\right) _{,a}=2C_{(1)ab}Q^{b}-L_{(1)b}A_{a}^{b}
\end{equation*}%
\begin{equation*}
L_{(1)a}=2C_{(0)ab}Q^{b}-L_{(0)b}A_{a}^{b}-G_{,a}.
\end{equation*}

- For $n=2$.

We have the QFI
\begin{eqnarray*}
I_{2} &=&\left( \frac{t^{2}}{2}C_{(2)ab}%
+tC_{(1)ab}+C_{(0)ab}\right) \dot{q}^{a}\dot{q}^{b}%
+t^{2}L_{(2)a}\dot{q}^{a}+tL_{(1)a}\dot{q}^{a}+L_{(0)a}\dot{%
q}^{a}+\frac{t^{3}}{3}s+\frac{t^{2}}{2}L_{(1)a}Q^{a}+ \\
&&+tL_{(0)a}Q^{a}+G(q)
\end{eqnarray*}%
where $C_{(2)ab}$ is a KT computed from the relation
\begin{equation*}
C_{(1)ab}=-L_{(0)(a;b)}-2C_{(0)c(a}A_{b)}^{c},\enskip %
C_{(2)ab}=-L_{(1)(a;b)}-2C_{(1)c(a}A_{b)}^{c}
\end{equation*}%
whereas the vector $L_{(2)a}$ and the quantity $G\ $are computed from the
relations%
\begin{equation*}
L_{(2)(a;b)}=-C_{(2)c(a}A_{b)}^{c},\enskip L_{(2)a}Q^{a}=s,\enskip\left(
L_{(1)b}Q^{b}\right) _{,a}=2C_{(2)ab}Q^{b}-2L_{(2)b}A_{a}^{b}
\end{equation*}%
\begin{equation*}
L_{(1)a}=2C_{(0)ab}Q^{b}-L_{(0)b}A_{a}^{b}-G_{,a},\enskip %
L_{(2)a}=C_{(1)ab}Q^{b}-\frac{1}{2}L_{(1)b}A_{a}^{b}-\frac{1}{2}\left(
L_{(0)b}Q^{b}\right) _{,a}.
\end{equation*}

\bigskip
In a similar manner we continue for higher values of $n.$

We observe that for all values of $n$ the KTs $C_{(N)ab}$, the vectors $L_{(N)a}$ and hence the conditions for $I_{n}$
can be written in terms of the triplet $%
\{G(q),L_{(0)a},C_{(0)ab}=KT\}$.

\section{Killing Tensors and collineations}

\label{section.3}

A symmetry of a geometric object  $A$ generated by the vector field $\mathbf{X}$ is
an equation of the form
\begin{equation}
\mathcal{L}_{\mathbf{X}}A=B \label{eq.col.0}
\end{equation}%
where $B$ is a tensor with the same number of indices and the same symmetries of indices as $A$. In a Riemannian space the symmetries of geometric objects which are defined in terms of the metric and its
derivatives are called collineations. The basic collineation is
\begin{equation}
\mathcal{L}_{\mathbf{X}}g_{ab}= 2X_{(a;b)} =2\psi(x) g_{ab} \label{eq.col.0a}
\end{equation}
where $g_{ab}$ is the metric tensor of the space and the vector $\mathbf{X}$ is called a conformal Killing vector
(CKV) with conformal factor $\psi$. If $\psi _{;ij}=0$ the CKV is said to be a special CKV (SCKV). If $\psi _{,i}=0$ and $\psi\neq0$, $\mathbf{X}$ is called a homothetic vector (HV); and if $\psi =0$
a KV. The next set of collineations are of the form
\begin{equation}
\mathcal{L}_{\mathbf{X}}\Gamma_{bc}^{a}= 2\delta_{(b}^{a} \phi_{,c)}. \label{eq.col.0b}
\end{equation}
In that case the vector $\mathbf{X}$ is called a projective collineation (PC) with projective factor $\phi$. If $\phi_{;ab}=0$ (i.e. $\phi_{,a}$ is a gradient KV) is called a special PC (SPC). If $\phi_{,c} =0$ is called an affine collineation (AC).

In general it holds the identity\footnote{K. Yano, ``The Theory of Lie Derivatives and its Applications'', North-Holland Publishing Co., Amsterdam (1957). \label{Yanob}}
\begin{equation}
\mathcal{L}_{\mathbf{X}}\Gamma^{a}_{bc}= X^{a}{}_{;bc} - R^{a}{}_{bcd}X^{d} \label{eq.col.1}
\end{equation}
which when $\mathbf{X}$ is a PC takes the form
\begin{equation}
X^{a}{}_{;bc} -2\delta_{(b}^{a} \phi_{,c)}= R^{a}{}_{bcd}X^{d}. \label{eq.col.2}
\end{equation}
A well-known result is that if $f,S$ define gradient KVs then the vector $fS_{,a}$ is an AC (since for any gradient KV $R_{abcd} S^{;d}=0$).

The HV and the KVs are ACs. An AC which is not generated from neither KVs nor the HV is called a proper AC.

A direct consequence of (\ref{eq.col.2}) is that a PC $\mathbf{X}$ defines the KT
\begin{equation}
C_{ab}= X_{(a;b)} -2\phi g_{ab}. \label{eq.col.3}
\end{equation}%
If $\mathbf{X}$ is an AC, then $X_{(a;b)}$ is a KT of order two. Therefore
from the $s$ (say) proper ACs $\eta^{a}$ of a space one constructs the $s$ KTs $C_{ab}=\eta_{\left( a;b\right) }.$

Concerning the KTs (\ref{eq.col.3}) they can be written in
terms of a vector field $L^{a}$ in the reducible form $C_{ab}=L_{(a;b)}$. In that case $L_{a}=X_{a}+M_{a}$ where $X^{a}$ is a PC with projective factor $\phi$ and $M_{a}$ is a CKV with conformal factor $-2\phi$ (because $M_{(a;b)} =-2\phi g_{ab}$ and from (\ref{eq.col.0a}) $M_{(a;b)} =\psi g_{ab}$).

Therefore if a Riemannian space admits $m$ CKVs $M^{a}$ and $m$ PCs $X^{a}$ such that $\psi(\mathbf{M})= -2\phi(\mathbf{X})$, we construct $m$ KTs of order two of the form $C_{ab}=L_{(a;b)}$ where $L_{a}=M_{a}+X_{a}.$

The maximum number of linearly independent KTs\footnote{T.Y. Thomas, Proc. N.A.S. \textbf{32}, 10 (1946). \label{Thomas1946}}$^{,}$\footnote{C.D. Collinson, J. Phys. A: Gen. Phys. \textbf{4}, 756 (1971). \label{Collinson1965}} of order 2 in a
Riemannian (or pseudo-Riemannian) manifold of dimension $n$ is $\frac{n(n+1)^{2}(n+2)}{12}$ and this is the
necessary and sufficient condition for the space to be maximal symmetric, or of constant curvature (see Refs. \ref{Thompson1986A}, \ref{Glass Garfinkle 2010},  \ref{Thomas1946} and \ref{Collinson1965}).

It is important to note that not all KTs of order 2 in a maximal symmetric space are reducible, that is of the form $C_{ab}=L_{(a;b)}$. For example in the cases of $E^{2}$ and $E^{3}$ (see sections \ref{sec.E2.geometry} and \ref{sec.KTE3})  the reducible KTs are subcases of more general non-reducible KTs.

Concerning the KTs of the form $C_{ab}=L_{(a;b)}$ defined on $V^{n}$ we have the following proposition.

\begin{proposition}
\label{prop1} In a space $V^{n}$ the vector fields of the form
\begin{equation}
L_{a}=c_{1I}S_{I,a}+c_{2A}M_{Aa}+c_{3}HV_{a}+ c_{4}AC_{a}+c_{5IJ}S_{I}S_{J,a}+2c_{6IA}S_{I}M_{Aa}+ c_{7K}(PC_{Ka} + CKV_{Ka})
\label{FL.20}
\end{equation}%
where $S_{I,a}$ are the gradient KVs, $M_{Aa}$ are the non-gradient KVs, $HV_{a}$ is the homothetic vector, $AC_{a}$ are the proper ACs, $S_{I}S_{J,a}$ are non-proper ACs, $PC_{Ka}$ are proper PCs with a projective factor $\phi_{K}$ and $CKV_{Ka}$ are conformal KVs with conformal factor $-2\phi_{K}$, produce the KTs of order 2 of the form $C_{ab}=L_{(a;b)}$. In the case of maximally symmetric spaces
there do not exist proper PCs and proper ACs\footnote{A. Barnes, Class. Quant. Grav. \textbf{10}(6), 1139 (1993). \label{Barnes}} therefore only the vectors
generated by the KVs are necessary, that is in these spaces
\begin{equation}
L_{a}=c_{1I}S_{I,a}+c_{2K}M_{Ka}+c_{3}HV_{a}+c_{5IJ}S_{I}S_{J,a}+ 2c_{6IK}S_{I}M_{Ka}.
\label{FL.21}
\end{equation}
The KVs alone give the solution $C_{ab}=0$ and the HV generates the trivial KT $g_{ab}$.
\end{proposition}

The special projective Lie algebra of a maximally symmetric space consists
of the vector fields of Table 1 $(I,J=1,2,...,n)$.
\bigskip

\begin{center}
\begin{tabular}{|l|l|l|}
\hline
\multicolumn{1}{|l|}{Collineation} & Gradient & Non-gradient \\ \hline
\multicolumn{1}{|l|}{Killing vectors (KV)} & $\mathbf{K}_{I}= \delta_{I}^{i} \partial_{i}$ & $\mathbf{X}_{IJ}=\delta _{\lbrack I}^{j}\delta
_{j]}^{i}x_{j}\partial _{i}$ \\ \hline
\multicolumn{1}{|l|}{Homothetic vector (HV)} & $\mathbf{H}=x^{i}\partial
_{i}$ &  \\ \hline
\multicolumn{1}{|l|}{Affine Collineations (AC)} & $\mathbf{A}%
_{II}=x_{I}\delta _{I}^{i}\partial _{i}$ & $\mathbf{A}_{IJ}=x_{J}\delta
_{I}^{i}\partial _{i}$, $I\neq J$ \\ \hline
\multicolumn{1}{|l|}{Special Projective collineations (SPC)} &  & $\mathbf{P}_{I}=x_{I}\mathbf{H}$ \\ \hline
\multicolumn{3}{l}{Table 1: Collineations of Euclidean space $E^{n}$.} \\
\end{tabular}
\end{center}
\bigskip

Therefore a maximally symmetric space of dimension $n$ admits

- $n$ gradient KVs and $\frac{n(n-1)}{2}$ non-gradient KVs

- 1 gradient HV

- $n^{2}$ non-proper ACs

- $n$ PCs which are special (that is the partial derivative of the projective function is a gradient KV).

\section{The KTs of $V^{2}$}

\label{sec.E2.geometry}

The KTs of the Euclidean space $E^{2}$ are well-known (see Ref. \ref{Tsamp2020}). However for the convenience of the reader we refer briefly the KTs of $V^{2}$  with metric $g_{ab}=(\varepsilon, 1)$ where $\varepsilon=\pm1$ in order  to include the 2d Minkowski space $L^{2}$.
\bigskip

- $V^{2}$ admits two gradient KVs $\partial _{x},\partial _{y}$ whose
generating functions are $x,y$ respectively and one non-gradient KV (the
rotation) $y\partial _{x}-\varepsilon x\partial _{y}$. These vectors can be written
collectively%
\begin{equation}
L^{a}=\left(
\begin{array}{c}
b_{1}+b_{3}y \\
b_{2}-\varepsilon b_{3}x%
\end{array}%
\right)  \label{FL.15}
\end{equation}%
where $b_{1},b_{2},b_{3}$ are arbitrary constants, possibly zero.

- The general KT of order 2 in $V^{2}$ is
\begin{equation}
C_{ab}=\left(
\begin{array}{cc}
\gamma y^{2}+2ay+A & -\gamma xy-ax-\beta y+C \\
-\gamma xy-ax-\beta y+C & \gamma x^{2}+2\beta x+B%
\end{array}%
\right).  \label{FL.14b}
\end{equation}%

- The vector $L^{a}$ generating KTs of $V^{2}$ of the form $%
C_{ab}=L_{(a;b)} $ is
\begin{equation}
L_{a}=\left(
\begin{array}{c}
-2\beta y^{2}+2axy+Ax+a_{8}y+a_{11} \\
-2ax^{2}+2\beta xy+a_{10}x+By+a_{9}%
\end{array}%
\right) .  \label{FL.14}
\end{equation}

- The KTs $C_{ab}=L_{(a;b)}$ in $V^{2}$ generated from the vector (\ref%
{FL.14}) are%
\begin{equation}
C_{ab}=L_{(a;b)}=\left(
\begin{array}{cc}
L_{x,x} & \frac{1}{2}(L_{x,y}+L_{y,x}) \\
\frac{1}{2}(L_{x,y}+L_{y,x}) & L_{y,y}%
\end{array}%
\right) =\left(
\begin{array}{cc}
2ay+A & -ax-\beta y+C \\
-ax-\beta y+C & 2\beta x+B%
\end{array}%
\right)  \label{FL.14.1}
\end{equation}%
where\footnote{%
Note that $L^{a}$ in (\ref{FL.14}) is the sum of the non-proper ACs of $%
E^{2} $ and not of its KVs which give $C_{ab}=0.$} $2C=a_{8}+a_{10}$.
Observe that these KTs are special cases of the general KTs (\ref{FL.14b})
for $\gamma =0$.

\section{The geometric quantities of $E^{3}$}

\label{sec.KTE3}

In $E^{3}$ the general KT of order 2 has independent components%
\begin{eqnarray}
C_{11} &=&\frac{a_{6}}{2}y^{2}+\frac{a_{1}}{2}%
z^{2}+a_{4}yz+a_{5}y+a_{2}z+a_{3}  \notag \\
C_{12} &=&\frac{a_{10}}{2}z^{2}-\frac{a_{6}}{2}xy-\frac{a_{4}}{2}xz-\frac{%
a_{14}}{2}yz-\frac{a_{5}}{2}x-\frac{a_{15}}{2}y+a_{16}z+a_{17}  \notag \\
C_{13} &=&\frac{a_{14}}{2}y^{2}-\frac{a_{4}}{2}xy-\frac{a_{1}}{2}xz-\frac{%
a_{10}}{2}yz-\frac{a_{2}}{2}x+a_{18}y-\frac{a_{11}}{2}z+a_{19}  \label{FL.E3}
\\
C_{22} &=&\frac{a_{6}}{2}x^{2}+\frac{a_{7}}{2}%
z^{2}+a_{14}xz+a_{15}x+a_{12}z+a_{13}  \notag \\
C_{23} &=&\frac{a_{4}}{2}x^{2}-\frac{a_{14}}{2}xy-\frac{a_{10}}{2}xz-\frac{%
a_{7}}{2}yz-(a_{16}+a_{18})x-\frac{a_{12}}{2}y-\frac{a_{8}}{2}z+a_{20}
\notag \\
C_{33} &=&\frac{a_{1}}{2}x^{2}+\frac{a_{7}}{2}%
y^{2}+a_{10}xy+a_{11}x+a_{8}y+a_{9}  \notag
\end{eqnarray}%
where $a_{I}$ with $I=1,2,...,20$ are arbitrary real constants.

The vector $L^{a}$ generating the KT $C_{ab}=L_{(a;b)}$ is
\begin{equation}
L_{a}=\left(
\begin{array}{c}
-a_{15}y^{2}-a_{11}z^{2}+a_{5}xy+a_{2}xz+2(a_{16}+a_{18})yz+a_{3}x
+2a_{4}y+2a_{1}z+a_{6} \\
-a_{5}x^{2}-a_{8}z^{2}+a_{15}xy-2a_{18}xz+a_{12}yz+
2(a_{17}-a_{4})x+a_{13}y+2a_{7}z+a_{14} \\
-a_{2}x^{2}-a_{12}y^{2}-2a_{16}xy+a_{11}xz+a_{8}yz+2(a_{19}-
a_{1})x+2(a_{20}-a_{7})y+a_{9}z+a_{10}%
\end{array}%
\right)  \label{eq.Kep.5}
\end{equation}
and the generated KT is
\begin{equation}
C_{ab}=\left(
\begin{array}{ccc}
a_{5}y+a_{2}z+a_{3} & -\frac{a_{5}}{2}x-\frac{a_{15}}{2}y+a_{16}z+a_{17} & -%
\frac{a_{2}}{2}x+a_{18}y-\frac{a_{11}}{2}z+a_{19} \\
-\frac{a_{5}}{2}x-\frac{a_{15}}{2}y+a_{16}z+a_{17} & a_{15}x+a_{12}z+a_{13}
& -(a_{16}+a_{18})x-\frac{a_{12}}{2}y-\frac{a_{8}}{2}z+a_{20} \\
-\frac{a_{2}}{2}x+a_{18}y-\frac{a_{11}}{2}z+a_{19} & -(a_{16}+a_{18})x-\frac{%
a_{12}}{2}y-\frac{a_{8}}{2}z+a_{20} & a_{11}x+a_{8}y+a_{9}%
\end{array}%
\right)  \label{eq.Kep.8}
\end{equation}%
which is a subcase of the general KT (\ref{FL.E3}) for $%
a_{1}=a_{4}=a_{6}=a_{7}=a_{10}=a_{14}=0$.

We note that the covariant expression of the most general KT $\Lambda _{ij}$
of order 2 of $E^{3}$ is\footnote{M. Crampin, Rep. Math. Phys. \textbf{20}, 31 (1984). \label{Crampin 1984}}$^{,}$\footnote{C. Chanu, L. Degiovanni and R.G. McLenaghan, J. Math. Phys. \textbf{47}, 073506 (2006). \label{Chanu 2006}}
\begin{equation}
\Lambda _{ij}=(\varepsilon _{ikm}\varepsilon _{jln}+\varepsilon
_{jkm}\varepsilon _{iln})A^{mn}q^{k}q^{l}+(B_{(i}^{l}\varepsilon
_{j)kl}+\lambda _{(i}\delta _{j)k}-\delta _{ij}\lambda _{k})q^{k}+D_{ij}
\label{CRA.46}
\end{equation}%
where $A^{mn},B_{i}^{l},D_{ij}$ are constant tensors all being symmetric and
$B_{i}^{l}$ also being traceless; $\lambda ^{k}$ is a constant vector. This
result is obtained from the solution of the Killing tensor equation in
Euclidean space.

Observe that $A^{mn}$, $D_{ij}$ have each 6 independent components; $%
B_{i}^{l}$ has 5 independent components; and $\lambda ^{k}$ has 3
independent components. Therefore $\Lambda _{ij}$ depends on $6+6+5+3=20$
arbitrary real constants, a result which is in accordance with the one found
earlier in (\ref{FL.E3}).

\section{Applications}
\label{applications}

In this section we discuss various applications of Theorem \ref{Theorem2}.

\subsection{Case of geodesics}

\label{sec.geodesics}

We apply Theorem \ref{Theorem2} to the geodesic equations in order to recover the results of Ref. \ref{Katzin 1981} in a simple and straightforward manner. In that case $Q^{a}=0$ and $A_{b}^{a}=0$; and the conditions of the FI \textbf{Integral 1} imply that $I_{n>2}=0$. Therefore the only QFI which survives is the
\begin{equation*}
I_{2}=\left( \frac{t^{2}}{2}G_{;ab}-tL_{(0)(a;b)}+C_{(0)ab}\right) \dot{q}%
^{a}\dot{q}^{b}-tG_{,a}\dot{q}^{a}+L_{(0)a}\dot{q}^{a}+G(q)
\end{equation*}%
where $C_{(0)ab}$, $G_{;ab}$ and $L_{(0)(a;b)}$ are KTs.

The FI $I_{2}$ consists of the three independent FIs\footnote{We ignore the index $(0)$ in order to simplify the notation.} (see Ref. \ref{Tsamp2020})
\begin{equation*}
I_{2a}=C_{ab}\dot{q}^{a}\dot{q}^{b},\enskip I_{2b}=\frac{t^{2}}{2}G_{;ab}%
\dot{q}^{a}\dot{q}^{b}-tG_{,a}\dot{q}^{a}+G(q),\enskip I_{2c}=-tL_{(a;b)}%
\dot{q}^{a}\dot{q}^{b}+L_{a}\dot{q}^{a}.
\end{equation*}
The time-dependent QFIs $I_{2b}$, $I_{2c}$ are the ones found in Ref. \ref{Katzin 1981}.  The QFI $I_{2a}$ is not found because the authors were looking only for time-dependent FIs.

\bigskip

\begin{tabular}{|l|l|}
\hline
QFI & Condition \\ \hline
$I_{2a} = C_{ab}\dot{q}^{a}\dot{q}^{b}$ & $C_{ab}=$ KT \\
$I_{2b} = \frac{t^{2}}{2} G_{;ab} \dot{q}^{a}\dot{q}^{b} - tG_{,a} \dot{q}^{a} + G(q)$ & $G_{;ab}=$ KT \\
$I_{2c} = -tL_{(a;b)}\dot{q}^{a}\dot{q}^{b} + L_{a}\dot{q}^{a}$ & $L_{(a;b)}= $ KT \\ \hline
\multicolumn{2}{l}{Table 2: The QFIs of geodesic equations.} \\
\end{tabular}

\bigskip

As an application of the above general results let us compute the FIs of the geodesic equations of the 3d metric\footnote{
If we set $z=it$,  the line element (\ref{eq.exa4.1}) takes the form
$ds^{2}= -dt^{2} -t^{2} \left( dx^{2} + dy^{2} \right)$
which is a conformally flat spacetime.}
\begin{equation}
ds^{2}=z^{2}\left( dx^{2}+dy^{2}\right) +dz^{2}. \label{eq.exa4.1}
\end{equation}

In this case the kinetic metric is  $g_{ab}=diag(z^{2},z^{2},1)$. The Ricci Scalar $R=-\frac{2}{z^{2}}$. Therefor this metric is not of constant curvature and consequently the number of KTs is less than 20.

The geodesic equations are
\begin{eqnarray}
\ddot{x} &=& -\frac{2}{z} \dot{x}\dot{z} \label{eq.exa4.2a} \\
\ddot{y} &=& -\frac{2}{z} \dot{y}\dot{z} \label{eq.exa4.2b} \\
\ddot{z} &=& z(\dot{x}^{2}+\dot{y}^{2}). \label{eq.exa4.2c}
\end{eqnarray}

Solving  the condition $C_{(ab;c)}=0$ we find that the metric (\ref{eq.exa4.1}) admits the following KTs
\begin{equation*}
C_{ab}=
\left(
  \begin{array}{ccc}
    \left( \frac{c_{1}}{z^{2}} +\frac{c_{2}}{2}y^{2} +c_{3}y +c_{4} \right) z^{4} & -\frac{1}{2}\left( c_{2}xy +c_{3}x +c_{5}y -2c_{7} \right)z^{4} & 0 \\
    -\frac{1}{2}\left( c_{2}xy +c_{3}x +c_{5}y -2c_{7} \right)z^{4} & \left( \frac{c_{1}}{z^{2}} +\frac{c_{2}}{2}x^{2} +c_{5}x +c_{6} \right)z^{4} & 0 \\
    0 & 0 & c_{1} \\
  \end{array}
\right)
\end{equation*}
where $c_{\kappa}$, $\kappa=1,2,...,7$ are arbitrary constants. Therefore there exist 7 linearly independent KTs as many as the free parameters involved.

In order to find the reducible KTs we solve the constraint $C_{ab}=L_{(a;b)}$ for a vector $L_{a}$. We have the following system of equations:
\begin{eqnarray*}
L_{1,1}+zL_{3} &=& c_{1}z^{2} +\frac{c_{2}}{2}y^{2}z^{4} +c_{3}yz^{4} +c_{4}z^{4} \\
L_{1,2} + L_{2,1} &=& -c_{2}xyz^{4} -c_{3}xz^{4} -c_{5}yz^{4} +2c_{7}z^{4} \\
zL_{1,3} + zL_{3,1} - 2L_{1} &=& 0 \\
L_{2,2}+zL_{3} &=& c_{1}z^{2} +\frac{c_{2}}{2}x^{2}z^{4} +c_{5}xz^{4} +c_{6}z^{4} \\
zL_{2,3} + zL_{3,2} -2L_{2} &=& 0 \\
L_{3,3} &=& c_{1}.
\end{eqnarray*}
The solution of the above system is the vector
\begin{equation*}
L_{a} =
\left(
  \begin{array}{c}
    z^{2}(b_{1}y+b_{2}) \\
    -z^{2}(b_{1}x+b_{3}) \\
    c_{1}z \\
  \end{array}
\right), \enskip
L_{(a;b)}= c_{1}\left(
             \begin{array}{ccc}
               z^{2} & 0 & 0 \\
               0 & z^{2} & 0 \\
               0 & 0 & 1 \\
             \end{array}
           \right)=c_{1}g_{ab}
\end{equation*}
that is $L_a$ is a homothetic vector.

In the case that the generating vector $L_{a}=G_{,a}$ we find that $b_{1}=b_{2}=b_{3}=0$ and $G=\frac{c_{1}}{2}z^{2}$. Then we have
\[
G_{,a}=
\left(
  \begin{array}{c}
    0 \\
    0 \\
    c_{1}z \\
  \end{array}
\right), \enskip
G_{;ab}= c_{1}\left(
             \begin{array}{ccc}
               z^{2} & 0 & 0 \\
               0 & z^{2} & 0 \\
               0 & 0 & 1 \\
             \end{array}
           \right) =c_{1}g_{ab}.
\]

In order to compute the QFIs for the geodesic equations of (\ref{eq.exa4.1}) we apply the results of Table 2. We have:
\bigskip

1) The QFI $I_{2a}$.
\begin{eqnarray*}
I_{2a} &=& C_{ab}\dot{q}^{a}\dot{q}^{b} \\
&=& \left( \frac{c_{1}}{z^{2}} + \frac{c_{2}}{2}y^{2} +c_{3}y +c_{4} \right)z^{4}\dot{x}^{2} - \left( c_{2}xy +c_{3}x +c_{5}y -2c_{7} \right)z^{4}\dot{x}\dot{y} + \left( \frac{c_{1}}{z^{2}} +\frac{c_{2}}{2}x^{2} +c_{5}x +c_{6} \right)z^{4} \dot{y}^{2} +c_{1}\dot{z}^{2} \\
&=& 2c_{1} \underbrace{\frac{1}{2} \left(z^{2}\dot{x}^{2} + z^{2}\dot{y}^{2} + \dot{z}^{2} \right)}_{=\text{kinetic energy}} - \frac{c_{2}}{2} z^{4} \left( x\dot{y} - y\dot{x} \right)^{2} + c_{3}z^{4}\dot{x} \left( x\dot{y} - y\dot{x} \right) + c_{4}z^{4}\dot{x}^{2} -c_{5}z^{4}\dot{y} \left( x\dot{y} - y\dot{x} \right) + \\
&& +c_{6}z^{4}\dot{y}^{2} + 2c_{7}z^{4}\dot{x}\dot{y}.
\end{eqnarray*}
This expression contains the independent FIs
\[
T= \frac{1}{2} \left(z^{2}\dot{x}^{2} + z^{2}\dot{y}^{2} + \dot{z}^{2} \right), \enskip I_{2a1}=z^{2}\dot{x}, \enskip I_{2a2}=z^{2}\dot{y}, \enskip I_{2a3}= z^{2}\left( x\dot{y} - y\dot{x} \right).
\]
We note that
\[
T = \frac{1}{2} \left( \dot{x}I_{2a1} + \dot{y}I_{2a2} + \dot{z}^{2} \right), \enskip I_{2a3}= xI_{2a2} - yI_{2a1}.
\]
\bigskip

2) The QFI $I_{2b}$.
\begin{eqnarray*}
I_{2b} &=& \frac{t^{2}}{2}G_{;ab}\dot{q}^{a}\dot{q}^{b} -tG_{,a}\dot{q}^{a} + G(q) \\
&=& c_{1}\frac{t^{2}}{2}\left( z^{2}\dot{x}^{2} + z^{2}\dot{y}^{2} + \dot{z}^{2} \right) -c_{1}tz\dot{z} + \frac{c_{1}}{2}z^{2}.
\end{eqnarray*}
Therefore
\[
I_{2b}= -t^{2}T +tz\dot{z} -\frac{z^{2}}{2}.
\]
\bigskip

3) The QFI $I_{2c}$.
\begin{eqnarray*}
I_{2c} &=& -tL_{(a;b)}\dot{q}^{a}\dot{q}^{b} + L_{a}\dot{q}^{a} \\
&=& -c_{1}t\left( z^{2}\dot{x}^{2} + z^{2}\dot{y}^{2} + \dot{z}^{2} \right) + z^{2}(b_{1}y+b_{2})\dot{x} -z^{2}(b_{1}x+b_{3})\dot{y} +c_{1}z\dot{z}
\end{eqnarray*}
which contains the new irreducible FI
\[
I_{2c1}=-tT + \frac{z\dot{z}}{2} = \frac{1}{2} \frac{d}{dt} \left(-t^{2}T + \frac{z^{2}}{2} \right).
\]
We note that
\[
I_{2b}= tI_{2c1} + \frac{tz\dot{z}}{2} - \frac{z^{2}}{2}.
\]

We collect the results in the following table.
\bigskip

\begin{tabular}{|l|}
\hline
$T= \frac{1}{2} \left(z^{2}\dot{x}^{2} + z^{2}\dot{y}^{2} + \dot{z}^{2} \right)$, \enskip $I_{2a1}=z^{2}\dot{x}$, \enskip $I_{2a2}=z^{2}\dot{y}$, \enskip $I_{2a3}= xI_{2a2} -yI_{2a1}$ \\
$I_{2c}=-tT + \frac{z\dot{z}}{2}$, \enskip $I_{2b}= -t^{2}T +tz\dot{z} -\frac{z^{2}}{2}$ \\
\hline
\multicolumn{1}{l}{Table 3: The LFIs/QFIs of geodesics of (\ref{eq.exa4.1}).} \\
\end{tabular}
\bigskip

Since the metric $g_{ab}$ is not flat the conjugate momenta $p_{a}$ of the Hamiltonian formalism are not equal to the velocities $\dot{q}^{a}$. Hence to compute the Poisson brackets (PBs) of the FIs we have to make the required transformation.

The conjugate momenta are
\[
p_{a}\equiv \frac{\partial T}{\partial \dot{q}^{a}} = g_{ab} \dot{q}^{b} =
\left(
  \begin{array}{c}
    z^{2}\dot{x} \\
    z^{2}\dot{y} \\
    \dot{z} \\
  \end{array}
\right).
\]
Then the FIs become
\bigskip

\begin{tabular}{|l|}
\hline
$T= \frac{1}{2} \left( \frac{p_{1}^{2}}{z^{2}} + \frac{p_{2}^{2}}{z^{2}} + p_{3}^{2} \right)$, \enskip $I_{2a1}=p_{1}$, \enskip $I_{2a2}=p_{2}$, \enskip $I_{2a3}= xp_{2} -yp_{1}$ \\
$I_{2c}=-tT + \frac{zp_{3}}{2}$, \enskip $I_{2b}= -t^{2}T +tzp_{3} -\frac{z^{2}}{2}$ \\
\hline
\multicolumn{1}{l}{Table 4: The LFIs/QFIs of geodesics of (\ref{eq.exa4.1}) in the phase space $(q^{a}, p_{a})$.} \\
\end{tabular}
\bigskip

We compute
\[
\{T, I_{2a1}\}= \{T, I_{2a2}\}= \{T, I_{2a3}\}= 0, \enskip \{T, I_{2b}\}= \frac{\partial I_{2b}}{\partial t}, \enskip \{T, I_{2c}\}= \frac{\partial I_{2c}}{\partial t}, \enskip \{I_{2a1}, I_{2a2}\}=0.
\]
The system is (Liouville) integrable because the three FIs $T, I_{2a1}, I_{2a2}$ are linearly independent and in involution.

Therefore we can find the solution of the system by quadrature using $T, I_{2a1}, I_{2a2}$. However, it is simpler to use instead of $T$ the time-dependent FI $I_{2c}$. Indeed we have
\[
\begin{cases}
z^{2}\dot{x} = k_{1} \\
z^{2}\dot{y} = k_{2} \\
2z\dot{z} = 2k_{4}t + k_{3}
\end{cases}
\implies
\frac{dz^{2}}{dt}= 2k_{4}t + k_{3}  \implies z(t)= \pm \left( k_{4}t^{2} + k_{3}t + k_{0} \right)^{1/2}
\]
where $k_{0}, k_{1}\equiv I_{2a1}, k_{2}\equiv I_{2a2}, k_{3}\equiv 4I_{2c}, k_{4}\equiv 2T$ are arbitrary constants.

Substituting in the remaining FIs we find
\[
\dot{x}=\frac{k_{1}}{z^{2}} \implies x(t)= \frac{2k_{1}}{(4k_{0}k_{4} - k_{3}^{2})^{1/2}} \tan^{-1} \left[ \frac{2k_{4}t + k_{3}}{(4k_{0}k_{4} - k_{3}^{2})^{1/2}} \right] +c
\]
and
\[
\dot{y}=\frac{k_{2}}{z^{2}} \implies y(t)= \frac{2k_{2}}{(4k_{0}k_{4} - k_{3}^{2})^{1/2}} \tan^{-1} \left[ \frac{2k_{4}t + k_{3}}{(4k_{0}k_{4} - k_{3}^{2})^{1/2}} \right] +c'
\]
where $c,c'$ are arbitrary constants.

The solution is
\[
q^{a}(t)=
\left(
  \begin{array}{c}
    x(t) \\
    y(t) \\
    z(t) \\
  \end{array}
\right) =
\left(
  \begin{array}{c}
    \frac{2k_{1}}{(4k_{0}k_{4} - k_{3}^{2})^{1/2}} \tan^{-1} \left[ \frac{2k_{4}t + k_{3}}{(4k_{0}k_{4} - k_{3}^{2})^{1/2}} \right] +c \\
    \frac{2k_{2}}{(4k_{0}k_{4} - k_{3}^{2})^{1/2}} \tan^{-1} \left[ \frac{2k_{4}t + k_{3}}{(4k_{0}k_{4} - k_{3}^{2})^{1/2}} \right] +c' \\
    \pm \left( k_{4}t^{2} + k_{3}t + k_{0} \right)^{1/2} \\
  \end{array}
\right).
\]

\subsection{The Whittaker dynamical system}

\label{sec.Whittaker}

The Whittaker dynamical system is a 2d Newtonian system with equations
\begin{equation*}
\ddot{x}=x,\enskip\ddot{y}=\dot{x}.
\end{equation*}
For that system the kinetic metric is the Euclidean metric $\delta_{ab}$ of $%
E^{2}$.

In the notation of the Theorem \ref{Theorem2} we have $A_{b}^{a}=\delta
_{2}^{a}\delta _{b}^{1}=\left(
\begin{array}{cc}
0 & 0 \\
1 & 0%
\end{array}%
\right) $ and $Q^{a}=V^{,a}=-x\delta _{1}^{a}=\left(
\begin{array}{c}
-x \\
0%
\end{array}%
\right)$ where $V=-\frac{1}{2}x^{2}$. \bigskip

We apply the Theorem \ref{Theorem2} to determine the QFIs. \bigskip

\textbf{Integral 1.}
\begin{eqnarray*}
I_{n} &=& \left( \frac{t^{n}}{n} C_{(n)ab} + ... + \frac{t^{2}}{2} C_{(2)ab}
+ t C_{(1)ab} + C_{(0)ab} \right) \dot{q}^{a} \dot{q}^{b} + t^{n} L_{(n)a}%
\dot{q}^{a} + ... + t^{2}L_{(2)a}\dot{q}^{a} + t L_{(1)a}\dot{q}^{a} +
L_{(0)a}\dot{q}^{a} + \\
&& + \frac{t^{n+1}}{n+1} L_{(n)a}Q^{a} + ... + \frac{t^{2}}{2} L_{(1)a}Q^{a}
+t L_{(0)a}Q^{a} + G(q)
\end{eqnarray*}
where $C_{(N)ab}$ are KTs. Taking into consideration the quantities mentioned above we find
\begin{equation*}
C_{(1)11} = -L_{(0)(1;1)} - 2C_{(0)12}, \enskip C_{(1)12} = -L_{(0)(1;2)} -
C_{(0)22}, \enskip C_{(1)22} = -L_{(0)(2;2)}
\end{equation*}
\begin{equation*}
C_{(k+1)11} = -L_{(k)(1;1)} - 2\frac{1}{k}C_{(k)12}, \enskip C_{(k+1)12} =
-L_{(k)(1;2)} - \frac{1}{k}C_{(k)22}, \enskip C_{(k+1)22} = -L_{(k)(2;2)}, %
\enskip k=1,...,n-1
\end{equation*}
\begin{equation*}
\left(-xL_{(k-1)1}\right)_{,1} = -2xC_{(k)11} - k(k+1)L_{(k+1)1} -kL_{(k)2}, %
\enskip \left(-xL_{(k-1)1}\right)_{,2} = -2xC_{(k)12} - k(k+1)L_{(k+1)2}, %
\enskip k=1,...,n-1
\end{equation*}
\begin{equation*}
L_{(n)(1;1)} =- \frac{2}{n}C_{(n)12}, \enskip L_{(n)(1;2)} = -\frac{1}{n}
C_{(n)22}, \enskip L_{(n)(2;2)} =0
\end{equation*}
\begin{equation*}
-xL_{(n)1}=s, \enskip \left(-xL_{(n-1)1}\right)_{,1} = -2xC_{(n)11} -
nL_{(n)2}, \enskip \left(-xL_{(n-1)1}\right)_{,2} = -2xC_{(n)12}
\end{equation*}
\begin{equation*}
G_{,1}= -2xC_{(0)11} - L_{(1)1} -L_{(0)2}, \enskip G_{,2}= -2xC_{(0)12} -
L_{(1)2}.
\end{equation*}

We note that all the QFIs $I_{n}(n>1)$ reduce to the QFI $I_{1}$. Therefore we continue only with the case $n=1$. We have
\begin{equation*}
I_{1}=\left( tD_{ab}+C_{ab}\right) \dot{q}^{a}\dot{q}^{b}+tL_{a}\dot{q}%
^{a}+B_{a}\dot{q}^{a}+\frac{t^{2}}{2}s+tB_{a}Q^{a}+G(q)
\end{equation*}%
where $C_{ab},D_{ab}$ are KTs and
\begin{equation}
D_{11}=-B_{(1;1)}-2C_{12},\enskip D_{12}=-B_{(1;2)}-C_{22},\enskip %
D_{22}=-B_{(2;2)}  \label{eq.wh1}
\end{equation}%
\begin{equation}
L_{(1;1)}=-2D_{12},\enskip L_{(1;2)}=-D_{22},\enskip L_{(2;2)}=0
\label{eq.wh2}
\end{equation}%
\begin{equation}
-xL_{1}=s  \label{eq.wh3}
\end{equation}%
\begin{equation}
\left( -xB_{1}\right) _{,1}=-2xD_{11}-L_{2},\enskip\left( -xB_{1}\right)
_{,2}=-2xD_{12}  \label{eq.wh4}
\end{equation}%
\begin{equation}
G_{,1}=-2xC_{11}-L_{1}-B_{2},\enskip G_{,2}=-2xC_{12}-L_{2}.  \label{eq.wh5}
\end{equation}

The KTs $C_{ab}$, $D_{ab}$ are of the form (see section \ref{sec.E2.geometry}%
)
\begin{equation*}
C_{ab}=\left(
\begin{array}{cc}
\gamma _{0}y^{2}+2a_{0}y+A_{0} & -\gamma _{0}xy-a_{0}x-\beta _{0}y+C_{0} \\
-\gamma _{0}xy-a_{0}x-\beta _{0}y+C_{0} & \gamma _{0}x^{2}+2\beta _{0}x+E_{0}%
\end{array}%
\right)
\end{equation*}%
and
\begin{equation*}
D_{ab}=\left(
\begin{array}{cc}
\gamma _{1}y^{2}+2a_{1}y+A_{1} & -\gamma _{1}xy-a_{1}x-\beta _{1}y+C_{1} \\
-\gamma _{1}xy-a_{1}x-\beta _{1}y+C_{1} & \gamma _{1}x^{2}+2\beta _{1}x+E_{1}%
\end{array}%
\right) .
\end{equation*}

Solving the conditions (\ref{eq.wh2}) we find
\begin{equation*}
L_{a}= \left(
\begin{array}{c}
\gamma_{1}x^{2}y+ a_{1}x^{2} + 2\beta_{1}xy - 2C_{1}x + k_{1} \\
-\gamma_{1}x^{3} - 3\beta_{1}x^{2} - 2E_{1}x + k_{2} \\
\end{array}
\right).
\end{equation*}
Substituting in (\ref{eq.wh3}) we get $a_{1}=\beta_{1}= \gamma_{1}= C_{}=
k_{1}=0$ and $s=0$. Therefore
\begin{equation*}
L_{a}= \left(
\begin{array}{c}
0 \\
-2E_{1}x + k_{2} \\
\end{array}
\right), \enskip D_{ab}= \left(
\begin{array}{cc}
A_{1} & 0 \\
0 & E_{1}%
\end{array}%
\right).
\end{equation*}
Solving the conditions (\ref{eq.wh1}) we find
\begin{equation*}
B_{a}= \left(
\begin{array}{c}
\gamma_{0}x^{2}y + 2\beta_{0}xy + a_{0}x^{2} - (A_{1}+2C_{0})x + k_{3} \\
-\gamma_{0}x^{3} - 3\beta_{0}x^{2} - 2E_{0}x - E_{1}y + k_{4} \\
\end{array}
\right)
\end{equation*}
which when replaced in (\ref{eq.wh4}) gives $a_{0}=\beta_{0}=\gamma_{0}=0$, $k_{2}=k_{3}$
and $E_{1}=2(A_{1}+C_{0})$. It follows:
\begin{equation*}
B_{a}= \left(
\begin{array}{c}
- (A_{1}+2C_{0})x + k_{2} \\
- 2E_{0}x - 2(C_{0}+A_{1})y + k_{4} \\
\end{array}
\right), \enskip C_{ab} = \left(
\begin{array}{cc}
A_{0} & C_{0} \\
C_{0} & E_{0}%
\end{array}%
\right).
\end{equation*}
Substituting in the integrability condition of (\ref{eq.wh5}) we find $%
A_{1}=0$ $\implies E_{1}=2C_{0}$. Therefore
\begin{equation*}
L_{a}= \left(
\begin{array}{c}
0 \\
-4C_{0}x + k_{2} \\
\end{array}
\right), \enskip D_{ab}= \left(
\begin{array}{cc}
0 & 0 \\
0 & 2C_{0}%
\end{array}%
\right), \enskip B_{a}= \left(
\begin{array}{c}
- 2C_{0}x + k_{2} \\
- 2E_{0}x - 2C_{0}y + k_{4} \\
\end{array}
\right), \enskip C_{ab} = \left(
\begin{array}{cc}
A_{0} & C_{0} \\
C_{0} & E_{0}%
\end{array}%
\right).
\end{equation*}
Finally integrating the conditions (\ref{eq.wh5}) we have
\begin{equation*}
G(x,y)= (E_{0}-A_{0})x^{2} + 2C_{0}xy - k_{4}x - k_{2}y.
\end{equation*}

The FI is
\begin{eqnarray*}
J_{1} &=& 2tC_{0}\dot{y}^{2} + A_{0}\dot{x}^{2} + 2C_{0}\dot{x} \dot{y} +
E_{0}\dot{y}^{2} - 4tC_{0}x\dot{y} + tk_{2}\dot{y} - 2C_{0}x\dot{x} + k_{2}%
\dot{x} - 2E_{0}x\dot{y} - 2C_{0}y\dot{y} + k_{4}\dot{y} + \\
&& + 2tC_{0}x^{2} - tk_{2}x + E_{0}x^{2}-A_{0}x^{2} + 2C_{0}xy - k_{4}x -k_{2}y
\end{eqnarray*}
which consists of the FIs
\begin{equation*}
J_{1a}= (\dot{y} - x) \left[ t(\dot{y}-x) + \dot{x}-y \right]
\end{equation*}
\begin{equation*}
J_{1b}= \dot{x}^{2} -x^{2}, \enskip J_{1c}= (\dot{y} - x)^{2}, \enskip %
J_{1d}= t(\dot{y}-x) + \dot{x} -y, \enskip J_{1e}= \dot{y}-x.
\end{equation*}

The independent FIs are the following:
\begin{equation*}
J_{11}= \dot{x}^{2}-x^{2}, \enskip J_{12}=\dot{y}-x, \enskip J_{13}= t(\dot{y%
}-x) + \dot{x} - y.
\end{equation*}
\bigskip

\textbf{Integral 2.}
\begin{equation*}
J_{2}= e^{\lambda t} \left( \lambda C_{ab} \dot{q}^{a} \dot{q}^{b} + \lambda
L_{a}\dot{q}^{a} + L_{a}Q^{a} \right)
\end{equation*}
where $\lambda \neq 0$, $C_{ab}$ is a KT, $\lambda C_{11} = - L_{(1;1)} -
2C_{12}$, $\lambda C_{12}= - L_{(1;2)} - C_{22}$, $\lambda C_{22}= -
L_{(2;2)}$ and $\left(-xL_{1}\right)_{,a} = -2\lambda xC_{1a} -
\lambda^{2}L_{a} - \lambda L_{2} \delta^{1}_{a}$.

We have the conditions
\begin{eqnarray}
L_{1,1} &=& - \lambda C_{11} - 2C_{12}  \label{Wh9} \\
L_{1,2} + L_{2,1} &=& - 2\lambda C_{12} - 2C_{22}  \label{Wh10} \\
L_{2,2} &=& - \lambda C_{22}  \label{Wh11} \\
\left(-xL_{1}\right)_{,a} &=& -2\lambda xC_{1a} - \lambda^{2} L_{a} -
\lambda L_{2} \delta^{1}_{a}.  \label{Wh12}
\end{eqnarray}

Solving the system of PDEs (\ref{Wh9})-(\ref{Wh11}) we find that
\begin{equation*}
L_{a}=\left(
\begin{array}{c}
ax^{2}+2\lambda \beta y^{2}+2(\beta -\lambda a)xy+k_{1}y-(\lambda
A+2C)x+k_{2} \\
(2\lambda a-3\beta )x^{2}-2\lambda \beta xy-\lambda By-(2\lambda
C+2B+k_{1})x+k_{3} \\
\end{array}%
\right)
\end{equation*}%
and
\begin{equation*}
C_{ab}=\left(
\begin{array}{cc}
2ay+A & -ax-\beta y+C \\
-ax-\beta y+C & 2\beta x+B%
\end{array}%
\right) .
\end{equation*}

Substituting in the last condition (\ref{Wh12}) we get $a=\beta =B=k_{3}=0.$
We consider the following subcases: \bigskip

i) Case $\lambda =\pm 1$.

We find $A=k_{1}=0$.

Then
\begin{equation*}
L_{a}=\left(
\begin{array}{c}
-2Cx+k_{2} \\
\mp 2Cx%
\end{array}%
\right) ,\enskip C_{ab}=C\left(
\begin{array}{cc}
0 & 1 \\
1 & 0%
\end{array}%
\right) .
\end{equation*}

The FI is
\begin{equation*}
J_{2a}= e^{\pm t} \left[ \pm 2C\dot{x}\dot{y} \pm (-2Cx + k_{2})\dot{x} - 2Cx%
\dot{y} + 2Cx^{2} - k_{2}x \right]
\end{equation*}
which contains the FIs
\begin{equation*}
J_{21} = e^{\pm t} (\dot{x} \mp x), \enskip J_{21b}= e^{\pm t} (\dot{y}-x)(%
\dot{x} \mp x) = J_{21} J_{12}.
\end{equation*}

We note that the FI $J_{11}$ can be derived from $J_{21+}$ and $J_{21-}$ as
follows
\begin{equation*}
J_{21+} J_{21-}= e^{t}(\dot{x} - x) e^{-t}(\dot{x}+x) = \dot{x}^{2} - x^{2}
= J_{11}.
\end{equation*}
Therefore $J_{11}$ is not an independent FI.

\bigskip

ii) Case $\lambda =\pm 2$.

We find $C=k_{1}=k_{2}=0$.

Then
\begin{equation*}
L_{a}=\left(
\begin{array}{c}
\mp 2Ax \\
0%
\end{array}%
\right) ,\enskip C_{ab}=A\left(
\begin{array}{cc}
1 & 0 \\
0 & 0%
\end{array}%
\right) .
\end{equation*}

We collect the results in the Table 5 below.

\subsubsection{Table of FIs}

\label{FIs of Whitakker system}

\begin{tabular}{|c|}
\hline
\multicolumn{1}{|l|}{$J_{12}=\dot{y}-x$} \\
\multicolumn{1}{|l|}{$J_{13}=t(\dot{y}-x)+\dot{x}-y=tJ_{12} +\dot{x}-y$} \\
\multicolumn{1}{|l|}{$J_{21\pm}=e^{\pm t}(\dot{x}\mp x)$} \\ \hline
\multicolumn{1}{l}{Table 5: FIs of Whittaker system.} \\
\end{tabular}
\bigskip

In order to study the integrability of the Whittaker system we compute the PBs of the independent FIs. We have
\[
\{J_{12},J_{13}\}=0, \enskip \{J_{12},J_{21\pm}\}= -e^{\pm t}, \enskip \{J_{13},J_{21\pm}\}= -e^{\pm t}(t\mp1), \enskip \{J_{21+},J_{21-}\}=-2.
\]
Therefore the 2d Whittaker system is integrable because the FIs $J_{12}$, $J_{13}$ are (functionally) independent and in involution.

However, the solution of the system can be found immediately by using $J_{12}$ and, instead of $J_{13}$, the time-dependent FIs $J_{21\pm}$. It follows that
\begin{equation}
x(t) = \frac{1}{2}(c_{-}e^{t} - c_{+}e^{-t}), \enskip y(t)= c_{0}t + \frac{1%
}{2}(c_{-}e^{t} + c_{+}e^{-t}) + c_{1}
\end{equation}
where $c_{\pm}$, $c_{0}$, $c_{1}$ are arbitrary constants.

\subsection{The autonomous linearly coupled 2d damped harmonic oscillator}

\label{sec.damped.oscillator}

This is the two-dimensional dynamical system with
equations of motion
\begin{eqnarray}
\ddot{x} + kx &=& py - 2m\dot{x}  \label{eq.Dj.1} \\
\ddot{y} + ky &=& - px - 2m\dot{y}  \label{eq.Dj.2}
\end{eqnarray}
where $m$, $p$, $k$ are (real or imaginary, non-zero) constants and $q^1=x$, $q^2=y$. The determination of the QFIs of this example have been discussed
before  (see example 6.5 in Ref. \ref{Djukic1975}) where it has been found one
new time-dependent QFI by giving arbitrary values to the quantities involved in the
weak Noether condition (equivalently the NBH equation). Using Theorem \ref%
{Theorem2} we shall recover this QFI plus a number of new QFIs not found
before.

A Lagrangian that describes this system is the Lagrangian of a 2d simple
harmonic oscillator
\begin{equation}
L=T-V=\frac{1}{2}\left( \dot{x}^{2}+\dot{y}^{2}\right) -\frac{1}{2}k\left(
x^{2}+y^{2}\right)  \label{eq.Dj.3}
\end{equation}%
with the generalized external forces
\begin{equation}
F^{a}=-P^{a}+A_{b}^{a}\dot{q}^{b}  \label{eq.Dj.4}
\end{equation}%
where $P^{a}=\left(
\begin{array}{c}
-py \\
px%
\end{array}%
\right) $ and $A_{b}^{a}=-2m\delta _{b}^{a}$. The equations of motion are written
\begin{equation}
\ddot{q}^{a}=-Q^{a}+A_{b}^{a}\dot{q}^{b}  \label{eq.Dj.5}
\end{equation}%
where $Q^{a}=V^{,a}+P^{a}=\left(
\begin{array}{c}
kx-py \\
ky+px%
\end{array}%
\right) $. The kinetic metric is the Euclidean metric $\delta_{ab}$ of the
plane $E^{2}$.

We apply Theorem \ref{Theorem2} to determine the QFIs of that system.
\bigskip

\textbf{Integral 1.}

The conditions of the QFI $I_{n}$ become
\begin{eqnarray}
C_{(1)ab} &=& -L_{(0)(a;b)} +4mC_{(0)ab}  \label{eq.dam1} \\
C_{(k+1)ab} &=& -L_{(k)(a;b)} + \frac{4m}{k}C_{(k)ab}, \enskip k=1,...,n-1
\label{eq.dam2} \\
\left(L_{(k-1)b} Q^{b} \right)_{,a} &=& 2C_{(k)ab}Q^{b} - k(k+1)L_{(k+1)a}
+2mkL_{(k)a}, \enskip k=1,...,n-1  \label{eq.dam3} \\
L_{(n)(a;b)} &=& \frac{4m}{n} C_{(n)ab}  \label{eq.dam4} \\
L_{(n)a}Q^{a}&=&s  \label{eq.dam5} \\
\left(L_{(n-1)b} Q^{b} \right)_{,a} &=& 2C_{(n)ab}Q^{b} +2mnL_{(n)a}
\label{eq.dam6} \\
G_{,a}&=& 2C_{(0)ab}Q^{b} - L_{(1)a} + 2mL_{(0)a}.  \label{eq.dam7}
\end{eqnarray}

Since $C_{(N)ab}=0$, $L_{(N)a}=0$ for $N=2,3,...,n$ it follows that only the QFI $I_{1}$ survives. Therefore\footnote{%
For simplicity we set $C_{(0)ab}\equiv C_{ab}$, $L_{(0)a}\equiv B_{a},$ $%
C_{(1)ab}\equiv D_{ab}$ and $L_{(1)a}\equiv L_{a}$.}
\begin{equation*}
I_{1}=\left( tD_{ab}+C_{ab}\right) \dot{q}^{a}\dot{q}^{b}+tL_{a}\dot{q}%
^{a}+B_{a}\dot{q}^{a}+\frac{t^{2}}{2}s+tB_{a}Q^{a}+G(q)
\end{equation*}%
where $C_{ab}=\frac{1}{4m}B_{(a;b)}+\frac{1}{16m^{2}}L_{(a;b)}$, $D_{ab}=%
\frac{1}{4m}L_{(a;b)}$ are KTs and
\begin{eqnarray}
L_{a}Q^{a} &=&s  \label{eq.DH1} \\
\left( B_{b}Q^{b}\right) _{,a} &=&2D_{ab}Q^{b}+2mL_{a}  \label{eq.DH2} \\
G_{,a} &=&2C_{ab}Q^{b}+2mB_{a}-L_{a}.  \label{eq.DH3}
\end{eqnarray}

Since $C_{ab}$, $D_{ab}$ are KTs we have that $B_{(a;b)}$, $L_{(a;b)}$ are
reducible KTs. Therefore from  section \ref{sec.E2.geometry} we have
\begin{equation*}
B_{a}=\left(
\begin{array}{c}
-2\beta_{1} y^{2}+2a_{1}xy+A_{1}x+n_{8}y+n_{11} \\
-2a_{1}x^{2}+2\beta_{1} xy+n_{10}x+B_{1}y+n_{9}%
\end{array}
\right), \enskip B_{(a;b)}= \left(
\begin{array}{cc}
2a_{1}y+A_{1} & -a_{1}x-\beta_{1} y+C_{1} \\
-a_{1}x-\beta_{1} y+C_{1} & 2\beta_{1} x+B_{1}%
\end{array}
\right)
\end{equation*}
\begin{equation*}
L_{a}=\left(
\begin{array}{c}
-2\beta_{2} y^{2}+2a_{2}xy+A_{2}x+w_{8}y+w_{11} \\
-2a_{2}x^{2}+2\beta_{2} xy+w_{10}x+B_{2}y+w_{9}%
\end{array}%
\right), \enskip L_{(a;b)}= \left(
\begin{array}{cc}
2a_{2}y+A_{2} & -a_{2}x-\beta_{2} y+C_{2} \\
-a_{2}x-\beta_{2} y+C_{2} & 2\beta_{2} x+B_{2}%
\end{array}
\right)
\end{equation*}
where $2C_{1}=n_{8}+n_{10}$ and $2C_{2}=w_{8}+w_{10}$.

Substituting $L_{a}$ in the condition (\ref{eq.DH1}) we obtain \underline{$%
a_{2}=\beta _{2}=s=0$} and there remain the following two cases:

1) $k=\pm ip$ with $w_{9}=\mp i w_{11}$, $w_{8}=\pm iB_{2},$ $w_{10}=\mp
iA_{2}$; and

2) $w_{9}=w_{11}=0$, $A_{2}=B_{2},$ $w_{8}= -w_{10}=\frac{k}{p}A_{2}$.
\bigskip

We continue the consideration of the remaining conditions for these two
cases.

1) Case $k=\pm ip$ with $w_{9}=\mp i w_{11}$, $w_{8}=\pm i B_{2}$ and $%
w_{10}=\mp i A_{2}$. \newline

1.1. The subcase $k=ip$.

Then
\begin{equation*}
L_{a}=\left(
\begin{array}{c}
A_{2}x+iB_{2}y+w_{11} \\
-iA_{2}x+B_{2}y-iw_{11}%
\end{array}%
\right) ,\enskip L_{(a;b)}=\left(
\begin{array}{cc}
A_{2} & \frac{i}{2}(B_{2}-A_{2}) \\
\frac{i}{2}(B_{2}-A_{2}) & B_{2}%
\end{array}%
\right)
\end{equation*}%
and the condition (\ref{eq.DH2}) gives
\begin{equation*}
a_{1}=\beta _{1}=0,\enskip(p-4im^{2})A_{2}=0,\enskip A_{2}=B_{2},\enskip %
w_{11}=0,\enskip n_{8}=iB_{1},\enskip n_{10}=-iA_{1},\enskip n_{9}=-in_{11}.
\end{equation*}%
Therefore
\begin{equation*}
B_{a}=\left(
\begin{array}{c}
A_{1}x+iB_{1}y+n_{11} \\
-iA_{1}x+B_{1}y-in_{11}%
\end{array}%
\right) ,\enskip B_{(a;b)}=\left(
\begin{array}{cc}
A_{1} & \frac{i}{2}(B_{1}-A_{1}) \\
\frac{i}{2}(B_{1}-A_{1}) & B_{1}%
\end{array}%
\right) .
\end{equation*}%
From $(p-4im^{2})A_{2}=0$ we have

1.1.1. Subcase $A_{2}=B_{2}=0$ $\implies L_{a}=0$, $D_{ab}=0$ and $C_{ab}=
\frac{1}{4m} B_{(a;b)}$.

From the integrability condition of (\ref{eq.DH3}) we find $(p-
4im^{2})(A_{1}+B_{1})=0$ which gives:

1.1.1.A. $A_{1}=-B_{1}$.

We have
\begin{equation*}
B_{a}=\left(
\begin{array}{c}
-B_{1}x+iB_{1}y+n_{11} \\
iB_{1}x+B_{1}y-in_{11}%
\end{array}%
\right) ,\enskip B_{(a;b)}=B_{1}\left(
\begin{array}{cc}
-1 & i \\
i & 1%
\end{array}%
\right) ,\enskip C_{ab}=\frac{B_{1}}{4m}\left(
\begin{array}{cc}
-1 & i \\
i & 1%
\end{array}%
\right)
\end{equation*}%
and integrating (\ref{eq.DH3}) we find
\begin{equation*}
G(x,y)=mB_{1}(-x^{2}+y^{2})+2imB_{1}xy+2mn_{11}(x-iy).
\end{equation*}

The FI is
\begin{eqnarray*}
J_{1}(11a) &=& -\frac{B_{1}}{4m} (\dot{x}^{2} - 2i\dot{x}\dot{y} -\dot{y}%
^{2}) + (-B_{1}x+iB_{1}y+n_{11})\dot{x} + (iB_{1}x+B_{1}y-in_{11})\dot{y} +
mB_{1}(-x^{2}+y^{2}) + \\
&& + 2imB_{1}xy + 2mn_{11}(x-iy)
\end{eqnarray*}
which consists of the independent FIs
\begin{eqnarray*}
J_{1a}(11a) &=& - \frac{1}{4m}(\dot{x} -i\dot{y})^{2} + (-x+iy) \dot{x} +
(ix +y)\dot{y} -m (x-iy)^{2} \equiv J_{11+} \\
J_{1b}(11a) &=& i\dot{x} +\dot{y} + 2m(ix+y) \equiv J_{12+}.
\end{eqnarray*}

1.1.1.B. For $p=4im^{2}$.

Integrating condition (\ref{eq.DH3}) we find
\begin{equation*}
G(x,y)=-miA_{1}xy+miB_{1}xy-\frac{m}{2}%
(B_{1}-A_{1})(x^{2}-y^{2})+2mn_{11}(x-iy).
\end{equation*}

The FI is
\begin{eqnarray*}
J_{1}(11b) &=& \frac{A_{1}}{4m}\dot{x}^{2}-\frac{i}{4m}(A_{1}-B_{1})\dot{x}%
\dot{y} +\frac{B_{1}}{4m}\dot{y}^{2} +(A_{1}x+iB_{1}y)\dot{x}+in_{9}\dot{x}%
+(-iA_{1}x+B_{1}y)\dot{y}+n_{9}\dot{y}+ \\
&&+\frac{1}{2} m(A_{1}-B_{1})(x^{2}-y^{2})-im(A_{1}-B_{1})xy+2imn_{9}(x-iy)
\end{eqnarray*}%
which consists of the irreducible FIs
\begin{eqnarray*}
J_{1a}(11b) &=&\frac{1}{4m}\dot{x}^{2}-\frac{i}{4m}\dot{x}\dot{y}+ x\dot{x}%
-ix\dot{y}+\frac{1}{2}m(x^{2}-y^{2})-imxy \\
J_{1b}(11b) &=&\frac{1}{4m}\dot{y}^{2}+\frac{i}{4m}\dot{x}\dot{y}+ iy\dot{x}%
+y\dot{y}-\frac{1}{2}m(x^{2}-y^{2})+imxy \\
J_{1c}(11b) &=&i\dot{x}+\dot{y}+2imx+2my.
\end{eqnarray*}

1.1.2. Subcase $p=4im^{2}$. We have
\begin{equation*}
L_{a} = A_{2}\left(
\begin{array}{c}
x+iy \\
-ix+y%
\end{array}%
\right), \enskip L_{(a;b)}= A_{2}\delta_{ab}, \enskip C_{ab}= \frac{1}{4m}
\left(
\begin{array}{cc}
A_{1}+\frac{A_{2}}{4m} & \frac{i}{2}(B_{1}-A_{1}) \\
\frac{i}{2}(B_{1}-A_{1}) & B_{1}+\frac{A_{2}}{4m}%
\end{array}
\right).
\end{equation*}

From the integrability condition of (\ref{eq.DH3}) we find $A_{2}=0$.
Therefore $L_{a}=0$ and $C_{ab}=\frac{1}{4m} B_{(a;b)}$.

We retrieve the FI $J_{1}(11b)$. \bigskip

1.2. The subcase $k=-ip$.

Working similarly for the case $k=-ip$ we find the FIs:

1.2.1. $A_{1}=-B_{1}$.
\begin{eqnarray*}
J_{1}(k=-ip) &=&-\frac{B_{1}}{4m}\dot{x}^{2}-\frac{iB_{1}}{2m}\dot{x}\dot{y}+%
\frac{B_{1}}{4m}\dot{y}^{2}-B_{1}(x+iy)\dot{x}-in_{9}\dot{x}+ B_{1}(-ix+y)%
\dot{y}+n_{9}\dot{y}- \\
&&-mB_{1}(x^{2}-y^{2})-2imB_{1}xy-2imn_{9}x+2mn_{9}y
\end{eqnarray*}%
which gives the irreducible FIs
\begin{eqnarray*}
J_{11-} &=&-\frac{1}{4m}\dot{x}^{2}-\frac{i}{2m}\dot{x}\dot{y}+\frac{1}{4m}%
\dot{y}^{2}-(x+iy)\dot{x}+(-ix+y)\dot{y}-m(x^{2}-y^{2})-2imxy \\
J_{12-} &=&-i\dot{x}+\dot{y}-2imx+2my.
\end{eqnarray*}

1.2.2. $p=-4im^{2}$.
\begin{eqnarray*}
J_{1}(k=-ip=-4m^{2}) &=&\frac{A_{1}}{4m}\dot{x}^{2}+\frac{i}{4m}(A_{1}-B_{1})%
\dot{x}\dot{y}+\frac{B_{1}}{4m}\dot{y}^{2}+ (A_{1}x-iB_{1}y)\dot{x}-in_{9}%
\dot{x}+(iA_{1}x+B_{1}y)\dot{y}+n_{9}\dot{y}+ \\
&&+\frac{1}{2}%
m(A_{1}-B_{1})(x^{2}-y^{2})+im(A_{1}-B_{1})xy-2imn_{9}x+2mn_{9}y
\end{eqnarray*}%
which consists of the irreducible FIs
\begin{eqnarray*}
J_{1a} &=&\frac{1}{4m}\dot{x}^{2}+\frac{i}{4m}\dot{x}\dot{y}+x\dot{x}+ix\dot{%
y}+\frac{1}{2}m(x^{2}-y^{2})+imxy \\
J_{1b} &=&\frac{1}{4m}\dot{y}^{2}-\frac{i}{4m}\dot{x}\dot{y}-iy\dot{x}+y\dot{%
y}-\frac{1}{2}m(x^{2}-y^{2})-imxy \\
J_{1c} &=&-i\dot{x}+\dot{y}-2imx+2my.
\end{eqnarray*}%\bigskip

We note that collectively the FIs $J_{11\pm}$ and $J_{12\pm}$ are written
\begin{eqnarray*}
J_{11\pm} &=& -\frac{1}{4m} (\dot{x} \mp i\dot{y})^{2} + (-x\pm iy)\dot{x}%
+(\pm ix+y)\dot{y} -m(x^{2}-y^{2})\pm 2imxy \\
J_{12\pm} &=& \dot{x} \mp i\dot{y} + 2m(x \mp iy).
\end{eqnarray*}

We observe that $J_{11\pm} = - \frac{1}{4m} (J_{12\pm})^{2}$ therefore the FIs $J_{11\pm}$ are not irreducible.

2) Case $w_{9}=w_{11}=0$, $A_{2}=B_{2}$ and $w_{8}=-w_{10}= \frac{k}{p}B_{2}$
$\implies C_{2}=0$.

We have
\begin{equation*}
L_{a}= B_{2} \left(
\begin{array}{c}
x+ \frac{k}{p}y \\
-\frac{k}{p}x +y \\
\end{array}
\right), \enskip L_{(a;b)} =B_{2} \delta_{ab}.
\end{equation*}

Then
\begin{eqnarray*}
4mC_{ab} &=& B_{(a;b)} + \frac{1}{4m}L_{(a;b)} \\
&=& \left(
\begin{array}{cc}
2a_{1}y+A_{1}+\frac{B_{2}}{4m} & -a_{1}x-\beta_{1} y+ C_{1} \\
-a_{1}x-\beta_{1} y+C_{1} & 2\beta_{1} x+B_{1}+\frac{B_{2}}{4m}%
\end{array}
\right)
\end{eqnarray*}
which in (\ref{eq.DH2}) gives $a_{1}=\beta_{1}=0$ and the following subcases:

2.1. $k=\pm ip$, $n_{11}=\pm in_{9}$, $n_{8}=\pm iB_{1}$, $n_{10}= \mp
iA_{1} $ and $B_{2}(p \mp 4im^{2})=0$.

This subcase gives again the FIs found in the case 1.

2.2. $n_{9}=n_{11}=0$, $n_{8}=-n_{10}= \frac{k}{p}B_{1} - \frac{B_{2}}{4mp}%
\left( k+4m^{2} \right)$, $A_{1}=B_{1}$ and $B_{2}(p^{2}-4m^{2}k)=0$.

We have
\begin{equation*}
B_{a}=\left(
\begin{array}{c}
B_{1}x+\frac{k}{p}B_{1}y - \frac{B_{2}}{4mp}\left( k+4m^{2} \right)y \\
-\frac{k}{p}B_{1}x + \frac{B_{2}}{4mp}\left( k+4m^{2} \right)x+B_{1}y%
\end{array}
\right), \enskip B_{(a;b)}= B_{1}\delta_{ab}, \enskip 4mC_{ab}= \left(B_{1}
+ \frac{B_{2}}{4m} \right) \delta_{ab}.
\end{equation*}

2.2.A. $B_{2}=0$.

We have $L_{a}=0$,
\begin{equation*}
B_{a}=\left(
\begin{array}{c}
B_{1}x+\frac{k}{p}B_{1}y \\
-\frac{k}{p}B_{1}x+B_{1}y%
\end{array}%
\right) ,\enskip B_{(a;b)}=B_{1}\delta _{ab},\enskip4mC_{ab}=B_{1}\delta
_{ab}.
\end{equation*}

The integrability condition of (\ref{eq.DH3}) implies that $p^{2}=4m^{2}k$
for non-trivial FIs and we compute
\begin{equation*}
G(x,y)= B_{1} \left( \frac{k}{4m} + m \right)(x^{2} + y^{2}).
\end{equation*}

The FI is
\begin{equation*}
J_{1}=\frac{1}{4m}(\dot{x}^{2}+\dot{y}^{2})+\left( x+\frac{k}{p}y\right)
\dot{x}+\left( y-\frac{k}{p}x\right) \dot{y}+\left( \frac{k}{4m}+m\right)
(x^{2}+y^{2})\implies
\end{equation*}%
\begin{equation*}
\bar{J}_{1}=\frac{p}{k}J_{1}=\frac{1}{4m}\frac{p}{k}(\dot{x}^{2}+\dot{y}%
^{2})+\left( \frac{p}{k}x+y\right) \dot{x}+\left( \frac{p}{k}y-x\right) \dot{%
y}+p\left( \frac{1}{4m}+\frac{m}{k}\right) (x^{2}+y^{2}).
\end{equation*}

2.2.B. $p^{2}=4m^{2}k$.

The integrability condition of (\ref{eq.DH3}) implies $p^{2}=-4m^{4}$ $%
\implies k=-m^{2}$, $p=\pm 2im^{2}$ and by integration we compute
\begin{equation*}
G(x,y)=\frac{3m}{4}B_{1}(x^{2}+y^{2})-\frac{9}{16}B_{2}(x^{2}+y^{2}).
\end{equation*}

The FI is written
\begin{eqnarray*}
J_{1}(2.2) &=& t \frac{B_{2}}{4m} (\dot{x}^{2} + \dot{y}^{2}) + \frac{B_{1}}{%
4m}(\dot{x}^{2} + \dot{y}^{2}) + \frac{B_{2}}{16m^{2}}(\dot{x}^{2} + \dot{y}%
^{2}) + tB_{2} \left( x \pm \frac{i}{2}y \right) \dot{x} + \\
&& + tB_{2} \left( \mp \frac{i}{2}x + y\right) \dot{y} + B_{1} \left( x \pm
\frac{i}{2}y \right) \dot{x} \pm B_{2}\frac{3i}{8m}y\dot{x} + B_{1} \left( y
\mp \frac{i}{2}x \right) \dot{y} \mp B_{2}\frac{3i}{8m}x\dot{y} + \\
&& + t \frac{3m}{4} B_{2}(x^{2} + y^{2})+ \frac{3m}{4}B_{1}(x^{2}+y^{2}) -
\frac{9}{16} B_{2} (x^{2}+y^{2})
\end{eqnarray*}
which consists of the irreducible FIs
\begin{eqnarray*}
J_{1a}(2.2) &=& t \frac{1}{4m} (\dot{x}^{2} + \dot{y}^{2}) + \frac{1}{16m^{2}%
}(\dot{x}^{2} + \dot{y}^{2}) + t \left( x \pm \frac{i}{2}y \right) \dot{x} +
t \left( \mp \frac{i}{2}x + y\right) \dot{y}\pm \\
&& \pm \frac{3i}{8m}(y\dot{x} - x\dot{y}) + t \frac{3m}{4} (x^{2} + y^{2})-
\frac{9}{16} (x^{2}+y^{2})
\end{eqnarray*}
\begin{eqnarray*}
J_{1b}(2.2) &=& \frac{1}{4m}(\dot{x}^{2} + \dot{y}^{2}) + \left( x \pm \frac{%
i}{2}y \right) \dot{x} + \left( y \mp \frac{i}{2}x \right) \dot{y} + \frac{3m%
}{4}(x^{2}+y^{2}).
\end{eqnarray*}

\bigskip

\textbf{Integral 2.}
\begin{equation*}
J_{2}= e^{\lambda t} \left( \lambda C_{ab} \dot{q}^{a} \dot{q}^{b} + \lambda
L_{a}\dot{q}^{a} + L_{a}Q^{a} \right)
\end{equation*}
where $\lambda \neq 0$ and
\begin{eqnarray}
L_{(a;b)} &=& (4m - \lambda) C_{ab}  \label{cond8a} \\
\left(L_{b}Q^{b}\right)_{,a} &=& 2\lambda C_{ab} Q^{b} + \lambda(2m-
\lambda)L_{a}.  \label{cond8b}
\end{eqnarray}

Since $C_{ab}$ is a KT condition (\ref{cond8a}) implies that $L_{(a;b)}$ is
a KT as well.

Consider the following cases: \bigskip

1) Case $\lambda =4m$.

From (\ref{cond8a}) we find that $L_{a}$ is a KV, i.e. $L_{a}=(b_{1}+b_{3}y)%
\partial _{x}+(b_{2}-b_{3}x)\partial _{y}$.

Then condition (\ref{cond8b}) becomes
\begin{equation}
\left( L_{b}Q^{b}\right) _{,a}=8mC_{ab}Q^{b}-8m^{2}L_{a}.  \label{cond8bb}
\end{equation}

Substituting the KV $L_{a}$ and the KT (\ref{FL.14b}) in (\ref{cond8bb}) we
have the following six subcases:

1.1. $k=\frac{p^{2}}{4m^{2}}$, $C=0$, $A=B$, $a=\beta=\gamma=0$, $b_{1}=
b_{2}= 0$ and $b_{3}=-\frac{p}{m}A$.

The FI is
\begin{equation*}
J_{2}\left( k=\frac{p^{2}}{4m^{2}} \right) = e^{4mt} \left[ \dot{x}^{2} +
\dot{y}^{2} - \frac{p}{m} (y\dot{x} - x\dot{y}) + \frac{p^{2}}{4m^{2}}
(x^{2}+y^{2}) \right].
\end{equation*}

1.2. $k=\frac{p^{2}}{4m^{2}}$, $C=0$, $A=B$, $a=\beta=\gamma=0$, $b_{3}=-%
\frac{p}{m}A$, $p=\pm i(k+8m^{2})$ and $b_{1}=\mp ib_{2}$.

Substituting $k=\frac{p^{2}}{4m^{2}}$ in $p=\pm i(k+8m^{2})$ we get a second
order equation wrt $p$ with solutions
\begin{equation*}
p_{1} = \mp 8im^{2} \implies k_{1}=-16m^{2}, \enskip p_{2}= \pm 4im^{2}
\implies k_{2}=-4m^{2}.
\end{equation*}

We have the following FI (all satisfy $k=\frac{p^{2}}{4m^{2}}$):
\begin{eqnarray*}
J_{2} &=& e^{4mt} \left[ 4mA(\dot{x}^{2} + \dot{y}^{2}) + 4m \left( \mp i
b_{2} - \frac{p}{m}Ay \right)\dot{x} + 4m \left( b_{2} + \frac{p}{m}Ax
\right) \dot{y} + \right. \\
&& + \left. \left( \mp i b_{2} - \frac{p}{m}Ay \right)(kx-py) + \left( b_{2}
+ \frac{p}{m}Ax \right)(ky +px) \right] \\
&=& e^{4mt} \left[ 4mA(\dot{x}^{2} + \dot{y}^{2}) + 4m \left( \mp i b_{2} -
\frac{p}{m}Ay \right)\dot{x} + 4m \left( b_{2} + \frac{p}{m}Ax \right) \dot{y%
} + \right. \\
&& + \left. A\frac{p^{2}}{m} (x^{2}+y^{2}) + b_{2} (px + ky \mp ikx \pm ipy) %
\right]
\end{eqnarray*}
which consists of the FIs $J_{2}\left( k=\frac{p^{2}}{4m^{2}} \right)$ found
earlier in the subcase 1.1 and the
\begin{equation*}
J_{21}= e^{4mt} \left( \mp 4mi\dot{x} + 4m\dot{y} + px + ky \mp ikx \pm ipy
\right).
\end{equation*}
Specifically we have
\begin{equation*}
J_{21}(k=-16m^{2}, p=\mp 8im^{2}) = J_{21}(k=-4m^{2}, p=\pm 4im^{2}) =
e^{4mt} \left( \mp i\dot{x} + \dot{y} \pm 2im x -2m y \right).
\end{equation*}

1.3. $C=0$, $A=B=0$, $a=\beta=\gamma=0$, $b_{3}=0$, $b_{1}=\mp b_{2}$ and $%
p=\pm i (k+8m^{2})$.

We find again the FI $J_{21}$ of the case 1.2. that is
\begin{equation*}
J_{2}(p=\pm i (k+8m^{2}))= e^{4mt} \left( \mp 4mi\dot{x} + 4m\dot{y} + px +
ky \mp ikx \pm ipy \right).
\end{equation*}

1.4. $k=\pm ip$: $L_{a}=0$ and $C_{ab}=B\left(
\begin{array}{cc}
-1 & \pm i \\
\pm i & 1%
\end{array}%
\right) $.

We have the FI
\begin{equation*}
J_{2}(k=\pm ip)= e^{4mt} ( \dot{x}^{2} - \dot{y}^{2} \mp 2i \dot{x}\dot{y} ).
\end{equation*}

1.5. $k=\pm ip$: $L_{a}=\mp ib_{2}\partial _{x}+b_{2}\partial _{y}$, $%
C_{ab}=B\left(
\begin{array}{cc}
-1 & \pm i \\
\pm i & 1%
\end{array}%
\right) $ and $p=\pm i(k+8m^{2})=\pm 4im^{2}$ which implies $k=-4m^{2}$.

We have the irreducible FIs $J_{2}(k=\pm ip)$ found in the subcase 1.4 and
the FI (already found)
\begin{equation*}
J_{2}(1.5)= e^{4mt} \left( \mp 4mi\dot{x} + 4m\dot{y} + ky + px \mp ikx \pm
ipy \right).
\end{equation*}

1.6. $k=\pm ip$: $A=B \pm 2iC$, $b_{3}= 4m (C \mp iB)$, $b_{1} = \mp ib_{2}$
and $p= \pm i (k+8m^{2})=\pm 4im^{2}$ which implies $k=-4m^{2}$. Then we
write the FI
\begin{eqnarray*}
J_{2}(1.6) &=& e^{4mt} \left[ (B \pm 2iC)\dot{x}^{2} + B \dot{y}^{2} + 2C%
\dot{x}\dot{y} \mp ib_{2} \dot{x} + 4m (C \mp iB)y \dot{x} + \right. \\
&& + \left. b_{2}\dot{y} - 4m(C \mp iB)x\dot{y} \pm 2m i b_{2} x - 2m b_{2}y
\mp 4m^{2}i (C \mp iB) (x^{2} + y^{2}) \right]
\end{eqnarray*}
which consists of the FIs
\begin{eqnarray*}
J_{21}(1.6) &=& e^{4mt} \left[ \dot{x}^{2} + \dot{y}^{2} \mp 4mi (y \dot{x}
- x\dot{y}) - 4m^{2} (x^{2} + y^{2}) \right] \\
J_{22}(1.6) &=& e^{4mt} \left[ \pm i \dot{x}^{2} + \dot{x}\dot{y} + 2m (y
\dot{x} - x\dot{y}) \mp 2m^{2}i (x^{2} + y^{2}) \right] \\
J_{23}(1.6) &=& e^{4mt} \left( \mp i \dot{x} + \dot{y} \pm 2m i x - 2m y
\right).
\end{eqnarray*}
\bigskip

2) Case $\lambda \neq 4m$.

Condition (\ref{cond8a}) gives the KT
\begin{equation}
C_{ab}=\frac{1}{4m-\lambda }L_{(a;b)}  \label{cond8c}
\end{equation}%
where the vector
\begin{equation*}
L_{a}=\left(
\begin{array}{c}
-2\beta y^{2}+2axy+Ax+a_{8}y+a_{11} \\
-2ax^{2}+2\beta xy+a_{10}x+By+a_{9}%
\end{array}%
\right)
\end{equation*}%
generates the reducible KT
\begin{equation*}
L_{(a;b)}=\left(
\begin{array}{cc}
2ay+A & -ax-\beta y+C \\
-ax-\beta y+C & 2\beta x+B%
\end{array}%
\right)
\end{equation*}%
where $2C=a_{8}+a_{10}$.

Substituting (\ref{cond8c}) in (\ref{cond8b}) we get
\begin{equation}
\left(L_{b}Q^{b}\right)_{,a} = \frac{2\lambda}{4m-\lambda} L_{(a;b)} Q^{b} +
\lambda(2m- \lambda)L_{a}  \label{cond8d}
\end{equation}
which implies that
\begin{eqnarray}
a(\lambda -3m) &=& 0  \label{intc1} \\
\beta (\lambda -3m) &=& 0  \label{intc2} \\
pa - \frac{3}{5}(m^{2}-k)\beta &=& 0  \label{intc3} \\
\frac{3}{5}(m^{2}-k)a + p\beta &=& 0  \label{intc4} \\
A + B - \frac{\lambda^{2}}{2p} (a_{8} - a_{10}) &=& 0  \label{intc5} \\
4p(\lambda - 2m) B + \frac{\lambda^{3}}{2} (3a_{10} - a_{8}) -
2m\lambda^{2}(a_{10} + 2a_{8}) + 2k\lambda(a_{8} + a_{10}) + 8m^{2}\lambda
a_{8} - 4km(a_{8}+a_{10}) &=& 0  \label{intc6} \\
\left[ \lambda^{3} - 6m\lambda^{2} + 4(2m^{2}+k)\lambda - 8km \right] A +
p\lambda a_{8} + p (3\lambda - 8m)a_{10} &=& 0  \label{intc7} \\
\left[ \lambda^{3} - 6m\lambda^{2} + 4(2m^{2}+k)\lambda - 8km \right] B - p
(3\lambda - 8m)a_{8} - p\lambda a_{10} &=& 0  \label{intc8} \\
pa_{9} + \frac{\lambda^{3} - 6m\lambda^{2} + (8m^{2}+k)\lambda -4km}{\lambda
-4m} a_{11} &=& 0  \label{intc9} \\
\frac{\lambda^{3} - 6m\lambda^{2} + (8m^{2}+k)\lambda -4km}{\lambda -4m}
a_{9} - p a_{11} &=& 0.  \label{intc10}
\end{eqnarray}

The set of the above conditions leads to three distinct QFIs because
conditions (\ref{intc1}) - (\ref{intc4}) concern only the parameters $a$, $%
\beta $; conditions (\ref{intc5}) - (\ref{intc8}) the parameters $A$, $B$, $%
a_{8}$, $a_{10}$ and conditions (\ref{intc9}) - (\ref{intc10}) only the
parameters $a_{9}$, $a_{11}$. Therefore when we write the final form of the
QFI this will consist of three independent FIs one for each set of
parameters.

The crucial parameter is the $\lambda .$ We consider two cases $\lambda \neq
2m$ and $\lambda =2m$ (where in both cases it is assumed that $\lambda\neq 4m$).

2.1. The subcase $\lambda \neq 2m$.

When $\lambda \neq 2m$ we have the following subcases for each set of
parameters considered above.\bigskip

2.1.1. Non-vanishing parameters $a$, $\beta $ and the rest equal to zero
(i.e. $A=B=a_{8}=a_{10}=0$, $a_{9}=a_{11}=0)$.

In this case we have only the conditions (\ref{intc1}) - (\ref{intc4}).

Only for $\lambda = 3m$ the parameters $a$, $\beta$ can be non-zero and thus
give a non-trivial FI, because then (\ref{intc1}), (\ref{intc2}) vanish
identically.

In that case the linear system of (\ref{intc3}) - (\ref{intc4}) has the non-zero
solution $\beta =\pm ia$ when $p=\pm i\frac{3}{5}(m^{2}-k)$ and we have for
this set of parameters the QFI
\begin{eqnarray*}
J_{2}(2.1.1) &=&e^{3mt}\left[ 3y\dot{x}^{2}\pm 3ix\dot{y}^{2}-3(x\pm iy)\dot{%
x}\dot{y}+3m(\mp iy^{2}+xy)\dot{x}+3m(-x^{2}\pm ixy)\dot{y}+\right. \\
&&\left. +(\mp iy^{2}+xy)(kx-py)+(-x^{2}\pm ixy)(ky+px)\right] .
\end{eqnarray*}

2.1.2. Non-vanishing parameters $A$, $B$, $a_{8}$, $a_{10}$ and all
remaining zero.

In this case we have the conditions (\ref{intc5}) - (\ref{intc8}).

From (\ref{intc5}), (\ref{intc6}) the parameters $A$, $B$ are expressed as
linear combinations of $a_{8}$, $a_{10}$ since $\lambda \neq 2m$.

These expressions $A(a_{8},a_{10})$, $B(a_{8},a_{10})$ when replaced in (\ref%
{intc7}) and (\ref{intc8}) respectively give a homogeneous linear system.
This system has non-vanishing solution of the form $a_{8}=Da_{10}$ only when
$D=1$. In that case $a_{8}=a_{10}$ with $p=\pm \frac{i}{4}(\lambda
^{2}-4m\lambda +4k)$ which implies that $A=-B=\mp ia_{8}$.

Then
\begin{equation*}
C_{ab}=\frac{a_{8}}{4m-\lambda }\left(
\begin{array}{cc}
\mp i & 1 \\
1 & \pm i \\
\end{array}%
\right) ,\enskip L_{a}=a_{8}\left(
\begin{array}{c}
\mp ix+y \\
x\pm iy \\
\end{array}%
\right)
\end{equation*}%
and the QFI for this set of parameters is
\begin{eqnarray*}
J_{2}(2.1.2) &=&e^{\lambda t}\left[ \frac{\lambda }{4m-\lambda }(\mp i\dot{x}%
^{2}\pm i\dot{y}^{2}+2\dot{x}\dot{y})+\lambda (\mp ix+y)\dot{x}+\lambda
(x\pm iy)\dot{y}+\right. \\
&&\left. +(p\mp ik)(x^{2}-y^{2})+2(k\pm ip)xy\right] .
\end{eqnarray*}

2.1.3. Non-vanishing parameters $a_{9}$, $a_{11}$ and all remaining
parameters zero.

In this case we have the homogeneous linear system (\ref{intc9}) - (\ref%
{intc10}) which has the non-zero solution $a_{11}=\mp ia_{9}$ for $p=\pm i%
\frac{\lambda ^{3}-6m\lambda ^{2}+(8m^{2}+k)\lambda -4km}{\lambda -4m}$.
Therefore $C_{ab}=0$ and $L_{a}= a_{9} (\mp i \partial_{x} + \partial_{y})$.

Since
\begin{equation*}
p=\pm i\frac{\lambda ^{3}-6m\lambda ^{2}+(8m^{2}+k)\lambda -4km}{\lambda -4m}%
=\pm i\frac{(\lambda -4m)(\lambda ^{2}-2\lambda m+k)}{\lambda -4m}=\pm
i(\lambda ^{2}-2\lambda m+k)
\end{equation*}%
we end up with the FI which we find in case 3) below. \bigskip

2.2. The subcase $\lambda =2m$.

In this case condition (\ref{cond8d}) becomes (since $\lambda \neq 4m$)
\begin{equation}
\left( L_{b}Q^{b}\right) _{,a}=2L_{(a;b)}Q^{b}.  \label{cond8e}
\end{equation}

From (\ref{cond8e}) we find that $a=\beta =0$, $A=-B$, $a_{8}=a_{10}\implies
$ $C=a_{8}$ and we end up with the system
\begin{equation*}
\begin{cases}
pa_{9}+ka_{11}=0 \\
ka_{9}-pa_{11}=0%
\end{cases}%
\end{equation*}%
which leads to two subcases: 2.2.1) $a_{9}=a_{11}=0$ and 2.2.2) $k=\pm ip$
and $a_{9}=\mp ia_{11}$. \bigskip

2.2.1. Subcase $a_{9}=a_{11}=0$.

We have
\begin{equation*}
C_{ab}=\frac{1}{2m}\left(
\begin{array}{cc}
A & a_{8} \\
a_{8} & -A%
\end{array}%
\right) ,\enskip L_{a}=\left(
\begin{array}{c}
Ax+a_{8}y \\
a_{8}x-Ay%
\end{array}%
\right) .
\end{equation*}

The FI is
\begin{eqnarray*}
J_{2}(\lambda =2m) &=&e^{2mt}\left[ A\dot{x}^{2}-A\dot{y}^{2}+2a_{8}\dot{x}%
\dot{y}+2mAx\dot{x}+2ma_{8}y\dot{x}+2ma_{8}x\dot{y}-2mAy\dot{y}+\right. \\
&&\left. +Ak(x^{2}-y^{2})-Apxy+a_{8}kxy-a_{8}py^{2}+a_{8}kxy+a_{8}px^{2}
-Apxy \right]
\end{eqnarray*}%
which consists of the irreducible FIs
\begin{eqnarray*}
J_{2a}(2.2.1) &=&e^{2mt}\left[ \dot{x}^{2}-\dot{y}^{2}+2m(x\dot{x}-y\dot{y}%
)+k(x^{2}-y^{2})-2pxy\right] \\
J_{2b}(2.2.1) &=&e^{2mt}\left[ \dot{x}\dot{y}+m(y\dot{x}+x\dot{y})+\frac{p}{2%
}(x^{2}-y^{2})+kxy\right] .
\end{eqnarray*}%
The FI $J_{2b}(2.2.1)$ is the one found in Ref. \ref{Djukic1975} (see equation (38) in Ref. \ref{Djukic1975}).

2.2.2. Subcase $a_{9}=\mp ia_{11}$.

In that case $k=\pm ip$,
\begin{equation*}
C_{ab}=\frac{1}{2m}\left(
\begin{array}{cc}
A & a_{8} \\
a_{8} & -A%
\end{array}%
\right) ,\enskip L_{a}=\left(
\begin{array}{c}
Ax+a_{8}y+a_{11} \\
a_{8}x-Ay\mp ia_{11}%
\end{array}%
\right) .
\end{equation*}

We find again the two FIs of the case 3.1 and the additional FI
\begin{equation*}
J_{2}(2.2.2)= e^{2mt} (\dot{x} \mp i\dot{y}).
\end{equation*}

We note that the FI of the case 1.4 can be derived from the above FI as
follows
\begin{equation*}
\left[ J_{2}(2.2.2) \right]^{2}= e^{4mt} (\dot{x} \mp i\dot{y})^{2} =
e^{4mt} ( \dot{x}^{2} - \dot{y}^{2} \mp 2i \dot{x}\dot{y})= J_{2}(1.4).
\end{equation*}
\bigskip

3) Case $C_{ab}=0$.

Condition (\ref{cond8a}) implies that $L_{a}=(b_{1}+b_{3}y)\partial
_{x}+(b_{2}-b_{3}x)\partial _{y}$ is a KV and the remaining condition (\ref%
{cond8b}) is written
\begin{equation}
\left( L_{b}Q^{b}\right) _{,a}=\lambda (2m-\lambda )L_{a}.  \label{cond8f}
\end{equation}

Substituting the KV $L_{a}$ in (\ref{cond8f}) we get $b_{3}=0$ and
non-vanishing values for $b_{1},b_{2}$ only when
\begin{equation*}
p=\pm i(\lambda^{2} - 2\lambda m + k) \implies
\begin{cases}
\lambda_{1}= m + \sqrt{m^{2} - k \mp ip} \\
\lambda_{2}= m - \sqrt{m^{2} - k \mp ip}.%
\end{cases}%
\end{equation*}
Then $b_{1} = \mp ib_{2}$.

Observe that since $p\neq 0$ also $\lambda^{2} - 2\lambda m + k \neq 0$.

The FI is
\begin{equation*}
J_{2(3)}=e^{\lambda t}(\mp i\lambda \dot{x}+\lambda \dot{y}\mp ikx\pm
ipy+px+ky),\quad p=\pm i(\lambda ^{2}-2\lambda m+k).
\end{equation*}%
We collect the above results in the following Table (see section \ref{FIs of
dampted SHO}).

\subsubsection{Table of FIs of  example \ref{sec.damped.oscillator}}

\label{FIs of dampted SHO}

\begin{tabular}{|l|l|}
\hline
Condition & LFI, QFI \\ \hline
- & $J_{2a}=e^{2mt}\left[ \dot{x}^{2}-\dot{y}^{2}+2m(x\dot{x}-y\dot{y}%
)+k(x^{2}-y^{2})-2pxy\right] $ \\
-, FI (38) in Ref. \ref{Djukic1975} & $J_{2b}=e^{2mt}\left[ \dot{x}\dot{y}+m(y%
\dot{x}+x\dot{y})+\frac{p}{2}(x^{2}-y^{2})+kxy\right] $ \\ \hline
$k=\pm ip$ & $J_{12}=\dot{x}\mp i\dot{y}+2m(x\mp iy)$ \\
& $J_{29}=e^{2mt}(\dot{x}\mp i\dot{y})$ \\
& $J_{2a}, J_{2b}$ \\ \hline
$p=\pm 4im^{2}$, $k=-4m^{2}$ & $J_{13}=\frac{1}{4m}\dot{x}^{2}\mp \frac{i}{4m%
}\dot{x}\dot{y}+x\dot{x}\mp ix\dot{y}+\frac{1}{2}m(x^{2}-y^{2})\mp imxy$ \\
& $J_{14}=\frac{1}{4m}\dot{y}^{2}\pm \frac{i}{4m}\dot{x}\dot{y}\pm iy\dot{x}%
+y\dot{y}-\frac{1}{2}m(x^{2}-y^{2})\pm imxy$ \\
& $J_{26}=e^{4mt}\left[ \pm i\dot{x}^{2}+\dot{x}\dot{y}+2m(y\dot{x}-x\dot{y}%
)\mp 2m^{2}i(x^{2}+y^{2})\right] $ \\
& $J_{2a}, J_{2b}, J_{12}, J_{29}, J_{11}, J_{21}, J_{23}$
\\ \hline
$p=\pm 2im^{2}$, $k=-m^{2}$ & \makecell[l]{$J_{15} = t \frac{1}{4m}
(\dot{x}^{2} + \dot{y}^{2}) + \frac{1}{16m^{2}}(\dot{x}^{2} + \dot{y}^{2}) +
t \left( x \pm \frac{i}{2}y \right) \dot{x} + t \left( \mp \frac{i}{2}x +
y\right) \dot{y}\pm $ \\ \qquad \enskip $\pm \frac{3i}{8m}(y\dot{x} -
x\dot{y}) + t \frac{3m}{4} (x^{2} + y^{2})- \frac{9}{16} (x^{2}+y^{2})$} \\
& $J_{16}=\frac{1}{4m}(\dot{x}^{2}+\dot{y}^{2})+\left( x\pm \frac{i}{2}%
y\right) \dot{x}+\left( y\mp \frac{i}{2}x\right) \dot{y}+\frac{3m}{4}%
(x^{2}+y^{2})$ \\
& $J_{2a}, J_{2b}$, $J_{11}$, $J_{21}$ \\ \hline
$k=\frac{p^{2}}{4m^{2}}$ & $J_{11}=\frac{1}{4m}(\dot{x}^{2}+\dot{y}%
^{2})+\left( x+\frac{k}{p}y\right) \dot{x}+\left( y-\frac{k}{p}x\right) \dot{%
y}+\left( \frac{k}{4m}+m\right) (x^{2}+y^{2})$ \\
& $J_{21}=e^{4mt}\left[ \dot{x}^{2}+\dot{y}^{2}-\frac{p}{m}(y\dot{x}-x\dot{y}%
)+\frac{p^{2}}{4m^{2}}(x^{2}+y^{2})\right] $ \\
& $J_{2a}, J_{2b}$ \\ \hline
$p=\pm i(k+8m^{2})$ & $J_{23}=e^{4mt}\left( \mp 4mi\dot{x}+4m\dot{y}%
+px+ky\mp ikx\pm ipy\right) $ \\
& $J_{2a}, J_{2b}$ \\ \hline
\makecell[l]{$p = \pm \frac{i}{4} (\lambda^{2} - 4 m \lambda + 4k)$, \\
$\lambda \neq 2m,4m$} & \makecell[l]{$J_{27}= e^{\lambda t} \left[
\frac{\lambda}{4m - \lambda} ( \mp i\dot{x}^{2} \pm i\dot{y}^{2} +
2\dot{x}\dot{y} ) + \lambda (\mp i x + y)\dot{x} + \lambda(x \pm iy) \dot{y}
+ \right.$ \\ \qquad \enskip $\left. + (p \mp ik)(x^{2} - y^{2})+ 2(k \pm
ip)xy \right]$} \\
& $J_{2a}, J_{2b}$ \\ \hline
$p=\pm i(\lambda ^{2}-2\lambda m+k)$ & $J_{28}=e^{\lambda t}(\mp i\lambda
\dot{x}+\lambda \dot{y}\mp ikx\pm ipy+px+ky)$ \\
& $J_{2a}, J_{2b}$ \\ \hline
$p=\pm i\frac{3}{5}(m^{2}-k)$ & \makecell[l]{$J_{24}= e^{3mt} \left[ 3y
\dot{x}^{2} \pm 3ix\dot{y}^{2} - 3(x \pm iy)\dot{x}\dot{y} + 3m (\mp i y^{2}
+ xy) \dot{x} + 3m (-x^{2} \pm ixy)\dot{y} + \right.$ \\ \qquad \enskip
$\left. + (\mp i y^{2} + xy)(kx-py) + (-x^{2} \pm ixy)(ky+px) \right]$} \\
& $J_{2a}, J_{2b}$ \\ \hline
\multicolumn{2}{l}{Table 6: FIs of the 2d harmonic oscillator with external forces.} \\
\end{tabular}

\bigskip

In the above Table some sets of conditions are a subset of other
more general conditions. In that case the FIs corresponding to that more
general conditions are also FIs for the special subset of these conditions;
however the opposite does not hold. For example:

- The $J_{2a}$, $J_{2b}$ are FIs for all values of $k$,$p$,$m$.

- The set of conditions ($k=-4m^{2}$, $p=\pm 4im^{2}$) gives $p=\mp ik$ $%
\implies k=\pm ip$ which means that the $J_{12}$, $J_{29}$ are FIs of that
set in addition to $J_{13}$, $J_{14}$, $J_{26}$. Observe that for that
special set of conditions $J_{13}-J_{14}= \frac{1}{4m} (J_{12})^{2}$ and $J_{11}=J_{13}+J_{14}$.
\bigskip

NOTE 1: The set of conditions $k=\pm ip$ with $p_{+}=i(k+8m^{2})$ and $%
p_{-}=-i(k+8m^{2})$ implies that $k=-4m^{2}$ and $p=\pm 4im^{2}$. \newline

NOTE 2: For $k=\frac{p^{2}}{4m^{2}}$, $p=\pm i(k+8m^{2})$ we find that $%
p_{+}= 4im^{2}, -8im^{2}$ and $p_{-}= -4im^{2}, 8im^{2}$. In that case the corresponding
FI $J_{23}$ reduces to
\begin{equation*}
J_{23}(k=p^{2}/4m^{2})=e^{4mt}\left( \mp i\dot{x}+\dot{y}\pm 2imx-2my\right).
\end{equation*}

\subsubsection{Discussing integrability}

- For arbitrary values of $k,p,m$.

For arbitrary values of $k,p,m$ the dynamical system admits two time-dependent QFIs the $J_{2a}$, $J_{2b}$. These
FIs are not in involution, that is, their PB does not vanish. Therefore an
arbitrary 2d harmonic oscillator with arbitrary external forces is not in
general integrable; in order to achieve  integrability we have to look for special values
of $k,p,m$ where more FIs are admitted.
\bigskip

- $k=\pm ip$.

The FIs $J_{12}$, $J_{29}$ are functionally independent and in involution
since $\{J_{12\pm },J_{29\pm }\}=0$. Therefore the system in that special
case is Liouville integrable and can be integrated by quadratures.

Indeed we have
\begin{equation}
\begin{cases}
\dot{x}\mp i\dot{y}+2m(x\mp iy)=c_{1\pm } \\
e^{2mt}(\dot{x}\mp i\dot{y})=c_{2\pm }%
\end{cases}%
\implies z(t)=-\frac{c_{2\pm }}{2m}e^{-2mt}+\frac{c_{1\pm }}{2m}
\label{eq.intdj.1}
\end{equation}%
where $z(t)\equiv x(t)\mp iy(t)$ and $c_{1\pm}$, $c_{2\pm}$ are arbitrary complex constants.
\bigskip

- $p= \pm 4im^{2}$, $k=-4m^{2}$.

These values satisfy the conditions $k=\pm ip$, $k=\frac{p^{2}}{4m^{2}}$ and
$p= \pm i(k + 8m^{2})$. Since $k=\pm ip$ it is straightforward that the system is integrable (previous case).

\section{Conclusions}

\label{section.5}

We discussed the relation between the collineations of the kinetic metric and the
existence of QFIs of the form (\ref{FL.5})   for autonomous holonomic dynamical systems of the form
 (\ref{FL.5.1}). We reviewed previous results in the literature
and derived a system of equations whose solution   derives all the QFIs admitted.
In particular  a given dynamical system defines  its own geometry via the
kinetic metric and  Theorem \ref{Theorem2} shows  that the  collineations of the kinetic
metric are sufficient  for the computation of the FIs. In a sense then a dynamical system is  `constrained' by the geometry it contains, because it is the
collineations of this geometry which provide the FIs for the dynamical system
and consequently specify its evolution.

The direct method of computing the QFIs from the condition $dI/dt=0$ is complementary to the standard method of Noether which uses  the standard
formal calculations required by the Lie approach. It appears that in  cases of low dimension the method discussed in this work would be more convenient due to the strong results of Differential Geometry in the field of collineations. However it cannot be compared with the generality of the Noether approach.
 Moreover it has been shown that the characterization of some
FIs as non-Noetherian is  meaningless because all FIs may be associated with a gauged
weak Noether symmetry.

It would be useful one  to extend the geometric analysis on the
symmetry conditions (\ref{eq.veldep5}) - (\ref{eq.veldep10}) in order to
understand in detail how geometry is related to the force-term $F^{a}.$ Such
an analysis will provide important information for the geometric properties
of integrable models and also a possible geometric classification for the
separable and non-separable Hamiltonian systems. An equally important step is the general solution of the system of
equations (\ref{eq.veldep5}) - (\ref{eq.veldep10}) which will provide all
QFIs of (\ref{FL.5.1}) of the form (\ref{FL.5}).

\section{Data Availability}

The data that supports the findings of this study are available within the article.

\appendix

\section{Proof of Theorem \ref{Theorem2}}

Recall that
\begin{equation*}
K_{ab}(t,q)= C_{(0)ab}(q) + C_{(1)ab}(q) t + C_{(2)ab}(q) \frac{t^{2}}{2} +
... + C_{(n)ab}(q) \frac{t^{n}}{n}
\end{equation*}%
and
\begin{equation*}
K_{a}(t,q)= L_{(0)a}(q) + L_{(1)a}(q)t + L_{(2)a}(q)t^{2} + ... +
L_{(m)a}(q) t^{m}.
\end{equation*}

We consider various cases\footnote{%
Equation (\ref{eq.veldep10}) is not necessary, because the integrability
condition $K_{,[ab]}=0$ does not intervene in the calculations. However, it
has been checked  that equation (\ref{eq.veldep10}) is always
satisfied identically from the solutions of the other equations
of the system.}. \bigskip

\underline{\textbf{I. Case $\mathbf{n=m}$}} (both $n$,$m$ finite) \bigskip

From equation (\ref{eq.veldep6}) we obtain
\begin{equation*}
C_{(1)ab} = -L_{(0)(a;b)} -2C_{(0)c(a}A_{b)}^{c}, \enskip C_{(k)ab} =
-L_{(k-1)(a;b)} - \frac{2}{k-1}C_{(k-1)c(a}A_{b)}^{c}, \enskip k=2,...,n, %
\enskip L_{(n)(a;b)} =-\frac{2}{n} C_{(n)c(a}A_{b)}^{c}.
\end{equation*}

The condition (\ref{eq.veldep9}) implies
\begin{equation*}
L_{(n)a}Q^{a}=s, \enskip \left( L_{(n-1)b} Q^{b} \right)_{,a} =
2C_{(n)ab}Q^{b} - nL_{(n)b}A^{b}_{a},
\end{equation*}
\begin{equation*}
\left( L_{(k-2)b} Q^{b} \right)_{,a} = 2C_{(k-1)ab}Q^{b} - k(k-1)L_{(k)a} -
(k-1)L_{(k-1)b}A^{b}_{a}, \enskip k=2,...,n.
\end{equation*}

The solution of (\ref{eq.veldep8}) is
\begin{equation*}
K= L_{(0)a}Q^{a}t + L_{(1)a}Q^{a}\frac{t^{2}}{2} + ... + L_{(n)a}Q^{a} \frac{%
t^{n+1}}{n+1} + G(q).
\end{equation*}

Substituting in (\ref{eq.veldep7}) we find
\begin{equation*}
G_{,a}= 2C_{(0)ab}Q^{b} - L_{(1)a} - L_{(0)b}A^{b}_{a}.
\end{equation*}

The FI is
\begin{eqnarray*}
I_{n} &=& \left( \frac{t^{n}}{n} C_{(n)ab} + ... + \frac{t^{2}}{2} C_{(2)ab}
+ t C_{(1)ab} + C_{(0)ab} \right) \dot{q}^{a} \dot{q}^{b} + t^{n} L_{(n)a}%
\dot{q}^{a} + ... + t^{2}L_{(2)a}\dot{q}^{a} + t L_{(1)a}\dot{q}^{a} +
L_{(0)a}\dot{q}^{a} + \\
&& + \frac{t^{n+1}}{n+1} L_{(n)a}Q^{a} + ... + \frac{t^{2}}{2} L_{(1)a}Q^{a}
+t L_{(0)a}Q^{a} + G(q)
\end{eqnarray*}
where $C_{(N)ab}$ are KTs, $C_{(1)ab} = -L_{(0)(a;b)} -2C_{(0)c(a}A_{b)}^{c}$%
, $C_{(k+1)ab} = -L_{(k)(a;b)} - \frac{2}{k}C_{(k)c(a}A_{b)}^{c}$ for $%
k=1,...,n-1$, $L_{(n)(a;b)} =-\frac{2}{n} C_{(n)c(a}A_{b)}^{c}$, $%
L_{(n)a}Q^{a}=s$, $\left( L_{(n-1)b} Q^{b} \right)_{,a} = 2C_{(n)ab}Q^{b} -
nL_{(n)b}A^{b}_{a}$, $\left( L_{(k-1)b} Q^{b} \right)_{,a} = 2C_{(k)ab}Q^{b}
- k(k+1)L_{(k+1)a} - kL_{(k)b}A^{b}_{a}$ for $k=1,...,n-1$ and $G_{,a}=
2C_{(0)ab}Q^{b} - L_{(1)a} - L_{(0)b}A^{b}_{a}$.

We note that $I_{0} < I_{1} < I_{2} < I_{3} < I_{4} < ...$, that is each QFI
$I_{k}$ is a subcase of the next QFI $I_{k+1}$ for all $k \in \mathbb{N}$.
Therefore we have only one independent QFI the $I_{n}$. The value of $n$ is
determined by the symmetries of the kinetic metric and the dynamics of each
specific system.

Observe that for $A^{a}_{b}=0$ the FI $I_{n}$ reduces to
\begin{eqnarray*}
I_{ns} &=& \left( - \frac{t^{n}}{n} L_{(n-1)(a;b)} - ... - \frac{t^{2}}{2}
L_{(1)(a;b)} - t L_{(0)(a;b)} + C_{(0)ab} \right) \dot{q}^{a} \dot{q}^{b} +
t^{n} L_{(n)a}\dot{q}^{a} + ... + t^{2}L_{(2)a}\dot{q}^{a} + \\
&& + t L_{(1)a}\dot{q}^{a} + L_{(0)a}\dot{q}^{a}+ \frac{t^{n+1}}{n+1}
L_{(n)a}Q^{a} + ... + \frac{t^{2}}{2} L_{(1)a}Q^{a} +t L_{(0)a}Q^{a} + G(q)
\end{eqnarray*}
where $C_{(0)ab}$, $L_{(N)(a;b)}$ are KTs, $L_{(n)a}$ is a KV, $%
L_{(n)a}Q^{a}=s$, $\left( L_{(n-1)b} Q^{b} \right)_{,a} =
-2L_{(n-1)(a;b)}Q^{b}$, $\left( L_{(k-1)b} Q^{b} \right)_{,a} =
-2L_{(k-1)(a;b)}Q^{b} - k(k+1)L_{(k+1)a}$ for $k=1,...,n-1$ and $G_{,a}=
2C_{(0)ab}Q^{b} - L_{(1)a}$.

We shall prove that $I_{n}(A^{a}_{b}=0)$ consists of two independent FIs.

In the case that $A^{a}_{b}=0$ we have the following:

- For $n=0$.
\begin{equation*}
I_{0} = C_{(0)ab} \dot{q}^a \dot{q}^b + L_{(0)a} \dot{q}^a + st + G(q)
\end{equation*}
where $C_{(0)ab}$ is a KT, $L_{(0)a}$ is a KV, $L_{(0)a} Q^{a}=s$ and $%
G_{,a} = 2C_{(0)ab} Q^{b}$.

This FI consists of the independent FIs
\begin{eqnarray*}
I_{01} &=& C_{(0)ab} \dot{q}^a \dot{q}^b + G(q) \\
I_{02} &=& L_{(0)a} \dot{q}^a + st.
\end{eqnarray*}

- For $n=1$.
\begin{eqnarray*}
I_{1}&=& \left( -tL_{(0)(a;b)} + C_{(0)ab} \right) \dot{q}^{a} \dot{q}^{b} +
tL_{(1)a}\dot{q}^{a} + L_{(0)a}\dot{q}^{a} + \frac{t^{2}}{2}s +
tL_{(0)a}Q^{a} + G(q)
\end{eqnarray*}
where $C_{(0)ab}$, $L_{(0)(a;b)}$ are KTs, $L_{(1)a}$ is a KV, $%
L_{(1)a}Q^{a}=s$, $\left(L_{(0)b}Q^{b} \right)_{,a} = - 2L_{(0)(a;b)}Q^{b}$
and $G_{,a} =2C_{(0)ab}Q^{b} - L_{(1)a}$.

This FI consists of
\begin{eqnarray*}
I_{11} &=& C_{(0)ab} \dot{q}^{a} \dot{q}^{b} + tL_{(1)a}\dot{q}^{a} + \frac{%
t^{2}}{2}s + G(q) \\
I_{12} &=& -tL_{(0)(a;b)} \dot{q}^{a} \dot{q}^{b} + L_{(0)a}\dot{q}^{a} +
tL_{(0)a}Q^{a}.
\end{eqnarray*}

- For $n=2$.
\begin{eqnarray*}
I_{2} &=& \left( - \frac{t^{2}}{2}L_{(1)(a;b)} -tL_{(0)(a;b)} + C_{(0)ab}
\right) \dot{q}^{a} \dot{q}^{b} + t^{2}L_{(2)a}\dot{q}^{a} + t L_{(1)a}\dot{q%
}^{a} + L_{(0)a}\dot{q}^{a} + \frac{t^{3}}{3}s + \frac{t^{2}}{2}
L_{(1)a}Q^{a} + \\
&& +t L_{(0)a}Q^{a} + G(q)
\end{eqnarray*}
where $C_{(0)ab}$, $L_{(M)(a;b)}$ for $M=0,1$ are KTs, $L_{(2)a}$ is a KV, $%
L_{(2)a}Q^{a}=s$, $\left( L_{(1)b} Q^{b} \right)_{,a} = -2L_{(1)(a;b)}Q^{b}$%
, $\left( L_{(0)b} Q^{b} \right)_{,a} = -2L_{(0)(a;b)}Q^{b} -2L_{(2)a}$ and $%
G_{,a}= 2C_{(0)ab}Q^{b} - L_{(1)a}$.

This FI consists of
\begin{eqnarray*}
I_{21} &=& \left( - \frac{t^{2}}{2}L_{(1)(a;b)} + C_{(0)ab} \right) \dot{q}%
^{a} \dot{q}^{b} + t L_{(1)a}\dot{q}^{a} + \frac{t^{2}}{2} L_{(1)a}Q^{a} +
G(q) \\
I_{22} &=& -tL_{(0)(a;b)} \dot{q}^{a} \dot{q}^{b} + t^{2}L_{(2)a}\dot{q}^{a}
+ L_{(0)a}\dot{q}^{a} + \frac{t^{3}}{3}s + +t L_{(0)a}Q^{a}.
\end{eqnarray*}

- For $n=3$.
\begin{eqnarray*}
I_{3} &=& \left( - \frac{t^{3}}{3} L_{(2)(a;b)} - \frac{t^{2}}{2}
L_{(1)(a;b)} - t L_{(0)(a;b)} + C_{(0)ab} \right) \dot{q}^{a} \dot{q}^{b} +
t^{3} L_{(3)a}\dot{q}^{a} + t^{2}L_{(2)a}\dot{q}^{a} + t L_{(1)a}\dot{q}^{a}
+ L_{(0)a}\dot{q}^{a} + \\
&& + \frac{t^{4}}{4} s + \frac{t^{3}}{3} L_{(2)a}Q^{a} + \frac{t^{2}}{2}
L_{(1)a}Q^{a} +t L_{(0)a}Q^{a} + G(q)
\end{eqnarray*}
where $C_{(0)ab}$, $L_{(M)(a;b)}$ for $M=0,1,2$ are KTs, $L_{(3)a}$ is a KV,
$L_{(3)a}Q^{a}=s$, $\left( L_{(2)b} Q^{b} \right)_{,a} = -2L_{(2)(a;b)}Q^{b}$%
, $\left( L_{(1)b} Q^{b} \right)_{,a} = -2L_{(1)(a;b)}Q^{b} - 6L_{(3)a}$, $%
\left( L_{(0)b} Q^{b} \right)_{,a} = -2L_{(0)(a;b)}Q^{b} - 2L_{(2)a}$ and $%
G_{,a}= 2C_{(0)ab}Q^{b} - L_{(1)a}$.

This FI consists of
\begin{eqnarray*}
I_{31} &=& \left( - \frac{t^{2}}{2} L_{(1)(a;b)} + C_{(0)ab} \right) \dot{q}%
^{a} \dot{q}^{b} + t^{3} L_{(3)a}\dot{q}^{a} + t L_{(1)a}\dot{q}^{a} + \frac{%
t^{4}}{4} s + \frac{t^{2}}{2} L_{(1)a}Q^{a} + G(q) \\
I_{32} &=& \left( - \frac{t^{3}}{3} L_{(2)(a;b)} - t L_{(0)(a;b)} \right)
\dot{q}^{a} \dot{q}^{b}+ t^{2}L_{(2)a}\dot{q}^{a} + L_{(0)a}\dot{q}^{a} +
\frac{t^{3}}{3} L_{(2)a}Q^{a} +t L_{(0)a}Q^{a}.
\end{eqnarray*}

- For $n=4$.
\begin{eqnarray*}
I_{4} &=& \left( -\frac{t^{4}}{4} L_{(3)(a;b)} - \frac{t^{3}}{3}
L_{(2)(a;b)} - \frac{t^{2}}{2}L_{(1)(a;b)} - t L_{(0)(a;b)} + C_{(0)ab}
\right) \dot{q}^{a} \dot{q}^{b} + t^{4} L_{(4)a}\dot{q}^{a} + t^{3} L_{(3)a}%
\dot{q}^{a} + t^{2}L_{(2)a}\dot{q}^{a} + \\
&& + t L_{(1)a}\dot{q}^{a} + L_{(0)a}\dot{q}^{a} + \frac{t^{5}}{5} s + \frac{%
t^{4}}{4} L_{(3)a}Q^{a} + \frac{t^{3}}{3} L_{(2)a}Q^{a} + \frac{t^{2}}{2}
L_{(1)a}Q^{a} +t L_{(0)a}Q^{a} + G(q)
\end{eqnarray*}
where $C_{(0)ab}$, $L_{(M)(a;b)}$ for $M=0,...n-1$ are KTs, $L_{(4)a}$ ia a
KV, $L_{(4)a}Q^{a}=s$, $\left( L_{(3)b} Q^{b} \right)_{,a} =
-2L_{(3)(a;b)}Q^{b}$, $\left( L_{(2)b} Q^{b} \right)_{,a} =
-2L_{(2)(a;b)}Q^{b} -12 L_{(4)a}$, $\left( L_{(1)b} Q^{b} \right)_{,a} =
-2L_{(1)(a;b)}Q^{b} - 6L_{(3)a}$, $\left( L_{(0)b} Q^{b} \right)_{,a} =
-2L_{(0)(a;b)}Q^{b} - 2L_{(2)a}$ and $G_{,a}= 2C_{(0)ab}Q^{b} - L_{(1)a}$.

This FI consists of
\begin{eqnarray*}
I_{41} &=& \left( -\frac{t^{4}}{4} L_{(3)(a;b)} - \frac{t^{2}}{2}%
L_{(1)(a;b)} + C_{(0)ab} \right) \dot{q}^{a} \dot{q}^{b} + t^{3} L_{(3)a}%
\dot{q}^{a} + t L_{(1)a}\dot{q}^{a} + \frac{t^{4}}{4} L_{(3)a}Q^{a} + \frac{%
t^{2}}{2} L_{(1)a}Q^{a}+ G(q) \\
I_{42} &=& \left( - \frac{t^{3}}{3} L_{(2)(a;b)} - t L_{(0)(a;b)} \right)
\dot{q}^{a} \dot{q}^{b} + t^{4} L_{(4)a}\dot{q}^{a} + t^{2}L_{(2)a}\dot{q}%
^{a} + L_{(0)a}\dot{q}^{a} + \frac{t^{5}}{5} s+ \frac{t^{3}}{3}
L_{(2)a}Q^{a} +t L_{(0)a}Q^{a}.
\end{eqnarray*}

-For $n=5$.
\begin{eqnarray*}
I_{5} &=& \left( -\frac{t^{5}}{5} L_{(4)(a;b)} - \frac{t^{4}}{4}
L_{(3)(a;b)} - \frac{t^{3}}{3}L_{(2)(a;b)} - \frac{t^{2}}{2} L_{(1)(a;b)} -
t L_{(0)(a;b)} + C_{(0)ab} \right) \dot{q}^{a} \dot{q}^{b} + t^{5} L_{(5)a}%
\dot{q}^{a} + t^{4} L_{(4)a}\dot{q}^{a} + \\
&& + t^{3} L_{(3)a}\dot{q}^{a} + t^{2}L_{(2)a}\dot{q}^{a}+ t L_{(1)a}\dot{q}%
^{a} + L_{(0)a}\dot{q}^{a} + \frac{t^{6}}{6} s+ \frac{t^{5}}{5}
L_{(4)a}Q^{a} + \frac{t^{4}}{4} L_{(3)a}Q^{a} + \frac{t^{3}}{3}
L_{(2)a}Q^{a} + \frac{t^{2}}{2} L_{(1)a}Q^{a} + \\
&& + t L_{(0)a}Q^{a} + G(q)
\end{eqnarray*}
where $C_{(0)ab}$, $L_{(M)(a;b)}$ for $M=0,...n-1$ are KTs, $L_{(5)a}$ is a
KV, $L_{(5)a}Q^{a}=s$, $\left( L_{(4)b} Q^{b} \right)_{,a} = -2L_{(4)(a;b)}
Q^{b}$, $\left( L_{(3)b} Q^{b} \right)_{,a} = -2L_{(3)(a;b)} Q^{b} -20
L_{(5)a}$, $\left( L_{(2)b} Q^{b} \right)_{,a} = -2L_{(2)(a;b)} Q^{b} -12
L_{(4)a}$, $\left( L_{(1)b} Q^{b} \right)_{,a} = -2L_{(1)(a;b)}Q^{b} -
6L_{(3)a}$, $\left( L_{(0)b} Q^{b} \right)_{,a} = -2L_{(0)(a;b)}Q^{b} -
2L_{(2)a}$ and $G_{,a}= 2C_{(0)ab}Q^{b} - L_{(1)a}$.

The FI consists of
\begin{eqnarray*}
I_{51} &=& \left( - \frac{t^{4}}{4} L_{(3)(a;b)} - \frac{t^{2}}{2}
L_{(1)(a;b)}+ C_{(0)ab} \right) \dot{q}^{a} \dot{q}^{b} + t^{5} L_{(5)a}\dot{%
q}^{a} + t^{3} L_{(3)a}\dot{q}^{a} + t L_{(1)a}\dot{q}^{a} + \frac{t^{6}}{6}
s+ \frac{t^{4}}{4} L_{(3)a}Q^{a} + \\
&& + \frac{t^{2}}{2} L_{(1)a}Q^{a} + G(q) \\
I_{52} &=& \left( -\frac{t^{5}}{5} L_{(4)(a;b)} - \frac{t^{3}}{3}%
L_{(2)(a;b)} - t L_{(0)(a;b)} \right) \dot{q}^{a} \dot{q}^{b} + t^{4}
L_{(4)a}\dot{q}^{a} + t^{2}L_{(2)a}\dot{q}^{a} + L_{(0)a}\dot{q}^{a} + \frac{%
t^{5}}{5} L_{(4)a}Q^{a} + \\
&& + \frac{t^{3}}{3} L_{(2)a}Q^{a} +t L_{(0)a}Q^{a}.
\end{eqnarray*}

If we continue in the same way, we prove that for $A^{a}_{b}=0$ the FI $%
I_{n} $ consists of the independent FIs:
\begin{eqnarray*}
I_{\ell1} &=& \left( - \frac{t^{2\ell}}{2\ell} L_{(2\ell-1)(a;b)} - ... -
\frac{t^{4}}{4} L_{(3)(a;b)} - \frac{t^{2}}{2} L_{(1)(a;b)} + C_{(0)ab}
\right) \dot{q}^{a} \dot{q}^{b} + t^{2\ell-1} L_{(2\ell-1)a}\dot{q}^{a} +
... + t^{3}L_{(3)a}\dot{q}^{a} + \\
&& + t L_{(1)a}\dot{q}^{a} + \frac{t^{2\ell}}{2\ell} L_{(2\ell-1)a}Q^{a} +
... + \frac{t^{4}}{4} L_{(3)a}Q^{a} + \frac{t^{2}}{2} L_{(1)a}Q^{a} + G(q)
\end{eqnarray*}
where $C_{(0)ab}$, $L_{(M)(a;b)}$ for $M=1,3,...,2\ell-1$ are KTs, $\left(
L_{(2\ell-1)b} Q^{b} \right)_{,a} = -2L_{(2\ell-1)(a;b)}Q^{b}$, $\left(
L_{(k-1)b} Q^{b} \right)_{,a} = -2L_{(k-1)(a;b)}Q^{b} - k(k+1)L_{(k+1)a}$
for $k=2,4,...,2\ell-2$ and $G_{,a}= 2C_{(0)ab}Q^{b} - L_{(1)a}$.

\begin{eqnarray*}
I_{\ell2} &=& \left( - \frac{t^{2\ell+1}}{2\ell+1} L_{(2\ell)(a;b)} - ... -
\frac{t^{3}}{3} L_{(2)(a;b)} - t L_{(0)(a;b)} \right) \dot{q}^{a} \dot{q}%
^{b} + t^{2\ell} L_{(2\ell)a}\dot{q}^{a} + ... + t^{2}L_{(2)a}\dot{q}^{a} +
\\
&& + L_{(0)a}\dot{q}^{a}+ \frac{t^{2\ell+1}}{2\ell+1} L_{(2\ell)a}Q^{a} +
... + \frac{t^{3}}{3} L_{(2)a}Q^{a} +t L_{(0)a}Q^{a}
\end{eqnarray*}
where $L_{(M)(a;b)}$ for $M=0,2,...,2\ell$ are KTs, $\left( L_{(2\ell)b}
Q^{b} \right)_{,a} = -2L_{(2\ell)(a;b)}Q^{b}$ and $\left( L_{(k-1)b} Q^{b}
\right)_{,a} = -2L_{(k-1)(a;b)}Q^{b} - k(k+1)L_{(k+1)a}$ for $%
k=1,3,...,2\ell-1$.

Observe that the set of the constraints of the FI $I_{n}(A^{a}_{b}=0)$ is
divided into one set involving the odd vectors $L_{(2k+1)a}$, the KT $%
C_{0ab} $ and the function $G(q)$; and a second set involving only the even
vectors $L_{(2k)a}$. This explains why in that case $I_{n}$ consists of two
independent FIs. \bigskip

\underline{\textbf{II. Case $\mathbf{n \neq m}$}.} ($n$ or $m$ may be
infinite) \vspace{12pt}

We find QFIs that are subcases of those found in \textbf{Case I} and \textbf{%
Case III}.

\bigskip

\underline{\textbf{III. Both $\mathbf{n}$, $\mathbf{m}$ are infinite.}}
\bigskip

In this case we consider the solution to have the form\footnote{%
To find a solution we consider $C_{(0)ab}=c_{0}C_{ab}$, $C_{(1)ab}=
c_{1}C_{ab}$, ..., $C_{(n)ab}= nc_{n}C_{ab}$, $L_{(0)a}= d_{0}L_{a}$, $%
L_{(1)a}=d_{1}L_{a}$, ...., $L_{(m)a}=d_{m}L_{a}$.}
\begin{equation*}
K_{ab}(t,q) = g(t)C_{ab}(q), \enskip K_{a}(t,q)= f(t)L_{a}(q)
\end{equation*}
where the functions $g(t),f(t)$ are analytic so that they may be represented
by polynomial functions as follows
\begin{equation*}  \label{eq.thm1}
g(t) = \sum^n_{k=0} c_k t^k = c_0 + c_1 t + ... + c_n t^n
\end{equation*}
\begin{equation*}  \label{eq.thm2}
f(t) = \sum^m_{k=0} d_k t^k = d_0 + d_1 t + ... + d_m t^m.
\end{equation*}
Only the following subcase give a new independent FI (this is the $J_{2}$ of
the Theorem \ref{Theorem2}). All the other subcases give trivial results
already analyzed in the previous cases. \bigskip

\underline{\textbf{Subcase $\mathbf{(g = e^{\lambda t}}$, $\mathbf{f =
e^{\mu t})}$.}} $\lambda \mu \neq 0$.

\begin{equation*}
\begin{cases}
\eqref{eq.veldep6} \implies \lambda e^{\lambda t} C_{ab} + e^{\mu t}
L_{(a;b)} + 2e^{\lambda t}C_{c(a}A^c_{b)} = 0 \\
\eqref{eq.veldep7} \implies - 2 e^{\lambda t} C_{ab} Q^{b} + \mu e^{\mu t}
L_a + K_{,a} + e^{\mu t}L_bA^b_a = 0 \\
\eqref{eq.veldep8} \implies K_{,t} = e^{\mu t} L_a Q^{a} \\
\eqref{eq.veldep9} \implies \mu^2 e^{\mu t} L_a + \mu e^{\mu t} L_bA^b_a
+e^{\mu t} \left( L_b Q^{b}\right)_{;a} - 2 \lambda e^{\lambda t} C_{ab}
Q^{b} = 0.%
\end{cases}%
\end{equation*}

We consider the following subcases.

a) \underline{For $\lambda \neq \mu$:}

From \eqref{eq.veldep6} we have that $C_{ab} = -\frac{2}{\lambda}
C_{c(a}A^c_{b)}$ and $L_a$ is a KV.

From \eqref{eq.veldep9} we find that $C_{ab}Q^{b}=0$ and $\mu^2 L_a + \mu
L_bA^b_a + \left( L_b Q^{b} \right)_{,a}= 0$.

The solution of \eqref{eq.veldep8} is
\begin{equation*}
K=\frac{1}{\mu }e^{\mu t}L_{a}Q^{a} +G(q)
\end{equation*}%
which when replaced in \eqref{eq.veldep7} gives $G_{,a}=0$, that is $%
G(q)=const \equiv0$.

The FI is
\begin{eqnarray*}
I_{e}(\lambda \neq \mu )&=&e^{\lambda t}C_{ab}\dot{q}^{a}\dot{q}^{b}+e^{\mu
t}L_{a}\dot{q}^{a}+\frac{1}{\mu }e^{\mu t}L_{a}Q^{a}
\end{eqnarray*}%
where $C_{ab}=-\frac{2}{\lambda }C_{c(a}A_{b)}^{c}$ is a KT such that $%
C_{ab}Q^{b}=0$ and $L_{a}=-\frac{1}{\mu ^{2}}\left( L_{b}Q^{b}\right) _{,a}-%
\frac{1}{\mu }L_{b}A_{a}^{b}$ is a KV.

We note that $I_{e}(\lambda \neq \mu)$ consists of the two independent FIs
\begin{equation*}
J_{2a} = e^{\lambda t}C_{ab}\dot{q}^{a}\dot{q}^{b}, \enskip J_{2b}= e^{\mu
t}L_{a}\dot{q}^{a}+\frac{1}{\mu }e^{\mu t}L_{a}Q^{a}.
\end{equation*}
The FIs $J_{2a}$, $J_{2b}$ are new.

\bigskip

b) \underline{For $\lambda = \mu$:}

From \eqref{eq.veldep6} we have that $C_{ab} = - \frac{1}{\lambda} L_{(a;b)}
- \frac{2}{\lambda} C_{c(a}A^c_{b)}$.

From \eqref{eq.veldep9} we find that $\lambda^2 L_a + \lambda L_bA^b_a +
\left( L_b Q^{b}\right)_{,a} - 2 \lambda C_{ab} Q^{b} = 0$.

The solution of \eqref{eq.veldep8} is
\begin{equation*}
K=\frac{1}{\lambda }e^{\lambda t}L_{a}Q^{a} +G(q)
\end{equation*}%
which when replaced in \eqref{eq.veldep7} gives $G(q)=const \equiv 0$.

The FI is
\begin{equation*}
I_{e}(\lambda =\mu )=e^{\lambda t}C_{ab}\dot{q}^{a}\dot{q}^{b}+e^{\lambda
t}L_{a}\dot{q}^{a}+ \frac{1}{\lambda }e^{\lambda t}L_{a}Q^{a} \equiv J_{2}
\end{equation*}%
where $C_{ab}=-\frac{1}{\lambda }L_{(a;b)}-\frac{2}{\lambda }%
C_{c(a}A_{b)}^{c}$ is a KT and the vector $L_{a}=-\frac{1}{\lambda ^{2}}%
\left( L_{b}Q^{b}\right) _{,a}-\frac{1}{\lambda }L_{b}A_{a}^{b}+\frac{2}{%
\lambda }C_{ab}Q^{b}$.

We note that the FIs $J_{2a}$, $J_{2b}$ found previously are subcases of the
new FI $J_{2}$. Specifically $J_{2a}= J_{2}(L_{a}=0)$ and $%
J_{2b}=J_{2}(C_{ab}=0)$. Therefore, the \textbf{Case III} leads to only one
independent FI the $J_{2}$.

\bigskip The above complete the proof of Theorem \ref{Theorem2}.

\theendnotes

\end{document}